\documentclass[10pt,a4paper]{article}
\usepackage[latin1]{inputenc}
\usepackage[T1]{fontenc}
\usepackage{cite}

\usepackage{array}
\newcommand{\PreserveBackslash}[1]{\let\temp=\\#1\let\\=\temp}

\usepackage{amsmath}
\usepackage{amsfonts}

\usepackage{pstricks}
\usepackage{pst-plot}
\usepackage{fancybox}
\usepackage[dvips]{graphicx}
\usepackage[dvips]{color}
\usepackage[figuresright]{rotating}
\usepackage{pifont}

\def\UNI-{UNI\/{\textbullet\nobreak}}
\newcommand{\unic}{\textsf{UNI\hspace{-0.5ex}
    \raisebox{0.15ex}{\scalebox{1.0}{$\bullet$}}\hspace{0.1ex}C}}

\DeclareMathOperator{\re}  {Re}
\DeclareMathOperator{\im}  {Im}

\DeclareMathOperator{\I}   {i}
\DeclareMathOperator{\mdot}{\mspace{-2.0mu} \cdot \mspace{-2.0mu}}

\newcommand{\N}{\mathbb N}                         
\newcommand{\R}{\mathbb R}                         
\newcommand{\C}{\mathbb C}                         
\newcommand{\J}{\mathbb J}                         

\newcommand{\difp}[2]{\frac{\partial #1}{\partial #2}}
\newcommand{\difd}[2]{\frac{d#1}{d#2}}

\newcommand{\e}{\mathrm{e}}

\newcommand{\Ec}{E^\mathrm{c}}
\newcommand{\Wc}{W^\mathrm{c}}

\newcommand{\bmu}{\boldsymbol{\mu}}
\newcommand{\bxi}{\boldsymbol{\xi}}
\newcommand{\bet}{\boldsymbol{\eta}}
\newcommand{\bzt}{\boldsymbol{\zeta}}
\newcommand{\bdl}{\boldsymbol{\delta}}

\newcommand{\BPhi}{\boldsymbol{\Phi}}
\newcommand{\BPsi}{\boldsymbol{\Psi}}

\newcommand{\trp}{\mathrm{T}}

\newcommand{\cc}   [1]{{\overline{#1}}}

\newcommand{\dotb} [1]{{\dot{\mathbf{#1}}}}

\newcommand{\abs}  [1]{\lvert#1\rvert}

\newcommand{\Mdot} [2]{\vec{#1}\mdot\vec{#2}}

\newcommand{\stab} [1]{\vec{#1}_\mathrm{s}}

\newcommand{\Dfxx} {\vec{F}_{\vec{x}\vec{x}}}
\newcommand{\Dfxxx}{\vec{F}_{\vec{x}\vec{x}\vec{x}}}
\newcommand{\Dfp}  {\vec{F}_{\mu}}
\newcommand{\Dfxp} {\vec{F}_{\vec{x}\mu}}

\renewcommand{\vec}[1]{\mathbf{#1}}

\newlength{\mlen}

\newcommand{\eqn}  [1]{\mbox{(\ref{#1})}}
\newcommand{\eqns} [2]{\mbox{(\ref{#1})~and~(\ref{#2})}}
\newcommand{\eqnto}[2]{\mbox{(\ref{#1})--(\ref{#2})}}
\newcommand{\Eqn}  [1]{\mbox{Equation~(\ref{#1})}}
\newcommand{\Eqns} [2]{\mbox{Equations~(\ref{#1})~and~(\ref{#2})}}

\newcommand{\fig}  [1]{\mbox{Fig.~\ref{#1}}}

\newcommand{\tab}  [1]{\mbox{Table~\ref{#1}}}

\newcommand{\tabto}[2]{\mbox{Tables~\ref{#1}--\ref{#2}}}

\newcommand{\sect} [1]{\mbox{Section~\ref{#1}}}
\newcommand{\sects}[2]{\mbox{Sections~\ref{#1}~and~\ref{#2}}}
\newcommand{\Sect} [1]{\mbox{Section~\ref{#1}}}

\makeatletter
\def\@eqnacr{{\ifnum0=`}\fi\@ifstar{\@yeqnacr}{\@yeqnacr}}
 
\def\@yeqnacr{\@ifnextchar [{\@xeqnacr}{\@xeqnacr[\z@]}}
 
\def\@xeqnacr[#1]{\ifnum0=`{\fi}\cr \noalign{\vskip\jot\vskip #1\relax}}
 
\def\eqalign{\null\,\vcenter\bgroup\openup1\jot \m@th \let\\=\@eqnacr
\ialign\bgroup\strut
\hfil$\displaystyle{##}$&$\displaystyle{{}##}$\hfil\crcr}
\def\endeqalign{\crcr\egroup\egroup\,}
\makeatother

\makeatletter
\def\fnum@figure{\bfseries%
                 \figurename~\thefigure%
                 \mdseries}
\renewenvironment{figure}
               {\@float{figure}}
               {\end@float}
\def\fnum@table{\bfseries%
                 \tablename~\thetable%
                 \mdseries}
\renewenvironment{table}
               {\@float{table}}
               {\end@float}
\makeatother

\begin{document}
\title{Systematic Derivation of Amplitude Equations\\ and Normal
  Forms for Dynamical Systems}

\author{M.\ Ipsen\footnotemark[2]\
   \and F.\ Hynne\footnotemark[3]\
   \and P.\ G.\ Sørensen\footnotemark[3]%
}
\maketitle

\renewcommand{\thefootnote}{\fnsymbol{footnote}}
\footnotetext[2]{%
  \unic, 
  Danish Computing Center for Research and Education. 
  DTU, 
  Building~304,
  DK-2800 \mbox{Lyngby},
  Denmark.
  }
\footnotetext[3]{%
  Department of Chemistry,
  University of Copenhagen,
  H.C.Ørsted Institute,
  Universi\-tets\-parken~5,
  DK-2100 Copenhagen,
  Denmark.
  }
\renewcommand{\thefootnote}{\arabic{footnote}}

\pagestyle{myheadings}
\thispagestyle{plain}
\begin{abstract} 
  We present a systematic approach to deriving normal forms
  and related amplitude equations for flows and discrete
  dynamics on the center manifold of a dynamical system at
  local bifurcations and unfoldings of these.  We derive a
  general, explicit recurrence relation that completely
  determines the amplitude equation and the associated
  transformation from amplitudes to physical space.  At any
  order, the relation provides explicit expressions for all
  the nonvanishing coefficients of the amplitude equation
  together with straightforward linear equations for the
  coefficients of the transformation.  The recurrence relation
  therefore provides all the machinery needed to solve a given
  physical problem in physical terms through an amplitude
  equation.  The new result applies to any local bifurcation
  of a flow or map for which all the critical eigenvalues are
  semisimple (\emph{i.e}.\ have Riesz index unity).  The
  method is an efficient and rigorous alternative to more
  intuitive approaches in terms of multiple time scales.  We
  illustrate the use of the method by deriving amplitude
  equations and associated transformations for the most common
  simple bifurcations in flows and iterated maps.  The results
  are expressed in tables in a form that can be immediately
  applied to specific problems.
\end{abstract}

\newpage
\bfseries
The behavior of physical, chemical, and biological systems
near the onset of dynamic instabilities has considerable
importance.  Here the dynamics can be described efficiently in
a low-dimensional space of ``amplitudes'' by using so-called
amplitude equations or normal forms.  Amplitude equations like
the complex Ginzburg-Landau equation have been widely used to
describe the qualitative behavior of such systems.  The paper
provides a general method for setting up all the necessary
equations to solve a given physical problem quantitatively in
terms of amplitude equations.

For example, chemical oscillations and waves can be described
by differential equations which near the onset of oscillation
can be approximated in terms of complex amplitudes.  This
reduces the state space from the (possibly high-dimensional)
concentration space to a complex plane.  Such reductions
usually result in immense savings in computational effort.  To
describe the dynamics in terms of amplitudes one must first of
all determine the coefficients of the amplitude equation.  For
a quantitative study and comparison with experiments, one must
also determine the transformation from amplitudes to the
actual concentrations of all the chemical species.  The paper
provides tables of formulas for calculating all relevant
coefficients.  For instabilities not covered by the tables,
the formulas can be calculated by a simple recurrence
relation, the main result of the paper.
\mdseries

\section{Introduction}
\label{sec:Intro}
The present work grew out of a need to extend the complex
Ginzburg-Landau equation for a chemical reaction-diffusion
problem~\cite{MIPhD97,Ipsen96}.  In the process it became
clear that a systematic and efficient method of deriving
amplitude equations and associated transformations for local
bifurcations is lacking.  We provide such a method for an
important class of bifurcations in the present paper.

Briefly, the idea of an amplitude equation for an oscillatory
system is to represent the state of the system through a
complex amplitude (real amplitude and phase) which varies much
more slowly in space and time than the original dynamical
variables.  A slow variation of an amplitude as compared with
much faster oscillations apply near a supercritical Hopf
bifurcation.  In fact, the concept of an amplitude can be
generalized to non-oscillatory modes near local bifurcations.

A large difference of characteristic times of variation of
amplitudes versus period of oscillation, is the basis of
traditional ``methods of multiple
times''~\cite{New74,Kuramoto,New92,Nic95} to derive amplitude
equations.  Unfortunately, such methods are somewhat intuitive
(as opposed to rigorous) and are not easy to generalize.  The
property utilized in amplitude equations is the critical
slowing down of modes associated with a bifurcation.  This
means that amplitude equations can be related to normal forms
describing the essential features of the motion of a state
point in the center manifold at a bifurcation.  In this paper,
we suggest a precise and general definition of amplitude
equations associated with local bifurcations of a flow or
iterated map and develop a systematic method of deriving them.
We emphasize that the present work does not require any use of
normal form theory.  In fact, the derivation of the
traditional normal form theorem appears naturally as part of
the theory developed in this paper.

We restrict the discussion to bifurcations associated with
spatially homogeneous systems and exclude \emph{e.g.}\ Turing
bifurcations.  Consequently, we shall treat systems defined by
ordinary differential equations or iterated maps.  The theory
can be generalized to spatially inhomogeneous systems
\emph{e.g.}\ with a linear diffusion operator.

The idea of the method proposed is the following.  Consider a
dynamical system at a local bifurcation.  The motion of the
state point in the $r$-dimensional center manifold $\Wc$ is
described by $r$ (generalized) amplitudes $z_j$ ($j = 1,
\dots, r$).  These are coordinates of a point {\bf z} in the
center subspace $\Ec$ with respect to a basis of eigenvectors
of the linearized vector field or map $\vec{J}$.  We assume
that $\vec{J}$ has $r$ linearly independent eigenvectors in
$\Ec$, but make no assumptions about the other modes.

For a flow, we derive a set of differential equations for the
amplitudes $z_j$, the amplitude equation, and a transformation
$\vec{h}(\vec{z})$ from $\Ec$ to $\Wc$ so that the image
$\vec{h}\bigl(\vec{z}(t)\bigl)$ describes the dynamical
evolution of the state point on $\Wc$ when {\bf z}($t$) is the
appropriate solution to the amplitude equation.

The transformation $\vec{h}(\vec{z})$ is determined by the
condition that as many coefficients of the amplitude equation
as possible must vanish, \emph{i.e.}\ it is a special type of
normal form transformation.  The non-vanishing coefficients of
the amplitude equation are determined together with the
transformation $\vec{h}(\vec{z})$, order by order in an
iterative process.  The method is very efficient --- in one
process, it determines the amplitude equation (in normal form)
and its unfolding as well as the center manifold.  Initially,
all that is required is to determine the non-hyperbolic
eigenvalues of the Jacobian $\vec{J}$ and the corresponding
right and left eigenvectors.  (Thus, it is not necessary to
transform $\vec{J}$ into Jordan block form.)  At each order of
iteration, the only non-trivial calculations are the solution
of a few systems of linear equations.  The center manifold is
then locally determined by the transformation
$\vec{h}(\vec{z})$ that immediately provides the physical
solution from that of the amplitude equation.  Other
representations of the geometry of the center manifold can
easily be obtained from $\vec{h}(\vec{z})$.  The method for
iterated maps (including Poincaré maps associated with
periodic solutions to flows) is similar to that just described
for flows.

The method reported here makes it straightforward to solve a
given physical (chemical, biological, $\dots$) problem in
terms of an amplitude description.  Once the general explicit
formulas have been derived for the type of bifurcation in
question, the coefficients of the amplitude equation and the
transformation $\vec{h}(\vec{z})$ can be calculated for the
particular problem directly in terms of the vector field (or
map).  So the amplitude equation for the particular physical
problem can be written down immediately, and any solution to
it in terms of amplitudes can then be expressed in terms of
real physical quantities through the transformation
$\vec{h}(\vec{z})$.  To illustrate the method, we derive
general formulas for the most common simple bifurcations for
flows as well as iterated maps.  A more challenging
application of the method to derive an amplitude equation for
an extension of the complex Ginzburg-Landau equation in the
context of fold-Hopf bifurcations in reaction-diffusion
problems will be published separately.

A method similar to the one developed here has been obtained
previously by Bruno \cite{Bruno95} (and see the references
therein) for the general case.  However, the restriction to
semisimple critical eigen\-values allows us to derive results
for any such bifurcation in particularly simple form.  These
results, a recurrence relation \eqn{eq:GSolveUnfold} with
\eqn{eq:PhiDefUnfold} for flows and \eqn{eq:DiscPhiDefUn} for
iterated maps, provide explicit expressions for the amplitude
equation (or normal form),
\eqns{eq:AmplUnfold}{eq:DiscAmplUnfold} respectively, and
explicit linear equations for the transformation from
amplitudes to physical space, \eqn{eq:LinearEqUn} and
\eqn{eq:LinearEqResMap} respectively.  It is entirely
straight\-forward to obtain all the nonvanishing coefficients
of the amplitude equation and the associated transformation,
order by order to any desired order, and to apply these to
specific physical problems.

Explicit results for a number of specific bifurcations,
applicable at the bifurcation points, can also be found in the
recent book by Kutznetsov~\cite{Kutz95}.  Other approaches are
described for example by {Guckenheimer \&
Holmes}~\cite{GuckHolmes}, {Coullet \& Spiegel}~\cite{CS83},
Elphick~\emph{et al.}~\cite{Elph87}, and by
Golubitsky~\emph{et al.}~\cite{Golu88}.  For a more physically
oriented discussion, see also Crawford~\cite{JDC91}.  Other
important techniques for obtaining information on the dynamics
in a vicinity of bifurcation points are perturbation methods
which have been reviewed in detail by Iooss \&
Joseph~\cite{IJ90} and Hassard \emph{et al.}~\cite{HKW81}.
These approaches generally provide series expansions of the
non-trivial steady or periodic solutions that emerge at the
bifurcation points.

The paper is organized as follows.  In \sect{sec:Flows}, we
develop the amplitude description for flows on the center
manifold. \Sect{sec:Unfold} extends the results to describe
the unfolding of the derived amplitude equation.  The
corresponding theory for iterated maps is very similar to that
for flows and is given in \sect{sec:Disc}.
\Sect{sec:GenExamp} illustrates the use of the general method
by deriving amplitude equations and transformation equations
for the most common simple bifurcations.  It presents the
results as convenient tables that can be used immediately for
any specific physical problem covered by the tables.  We
illustrate the use of the tables in
appendix~\ref{sec:Examples} by deriving amplitude equations
and transformations for some well-known systems.

Amplitudes as initially defined in this paper
(\sects{sec:Flows}{sec:Disc}) have simple geometric
significance, common to all bifurcations for flows and
iterated maps.  However, for oscillatory modes, it is
advantageous to introduce the concept of ``proper
amplitudes'', which are complex variables for which both
modulus and argument vary slowly.  These proper amplitudes
agree with accepted usage.  It is then possible to scale time
and amplitudes for all bifurcations so that the scaled
amplitude equations become independent of the distance from
the bifurcation point to lowest order --- we define proper
amplitudes to incorporate such scaling.  Scalings and proper
amplitudes are discussed in \sect{sec:Scalings}.

The physical interpretation of unfoldings of amplitude
equations is discussed in \sect{sec:UnInt} aiming at a more
intuitive understanding of their meaning.  \Sect{sec:PhysRela}
discusses some features of the solutions to amplitude
equations, transformed back to the original physical space,
\emph{i.e.}\ solutions in physical terms to physical problems.
Such back transformation is an important part of the solution
and it follows naturally from the present method.  In the
final \sect{sec:Summary}, we summarize and discuss the results
of this paper.

\section{Amplitude equations for flows}
\label{sec:Flows}
We start by considering a flow on~$\R^n$
\begin{equation}
  \dotb{x} = \vec{F}(\vec{x}) = 
  \vec{J} \mdot \vec{x} + \vec{f}(\vec{x}),
  \label{eq:Ode}
\end{equation}
with a stationary solution $\vec{x} = \vec{0}$.  Here, the
vector field on the right-hand side has been split into linear
and nonlinear parts $\Mdot{J}{x}$ and $\vec{f}(\vec{x})$.  We
assume that $\vec{J}$ has $r$ distinct right eigenvectors
$\vec{u}_j$ and left eigenvectors $\vec{u}_j^*$ corresponding
to $r$ critical eigenvalues $\lambda_j$ with zero real part
($j = 1,\dots,r$), chosen to satisfy the biorthonormality
conditions

\begin{equation}
  \vec{u}_i^* \mdot \vec{u}_j = \delta_{ij}, 
  \quad \text{for \ }
  i,j = 1,\dots,r,
  \label{eq:BiOrtho}
\end{equation}
where $\delta_{ij}$ is the usual Kronecker delta.  The $r$
eigenvectors span the $r$-dimensional center subspace $\Ec$.
As discussed in~\cite{GuckHolmes}, the center manifold $\Wc$
can be locally parameterized by points $\vec{z} \in \Ec$
through a smooth transformation \mbox{$\vec{h}: \Ec
\rightarrow \R^n$:}
\begin{equation}
  \Wc = 
  \bigl\{ 
  \vec{x} \;\bigl|\;
  \vec{x} = \vec{z} + \vec{h}(\vec{z}) 
  \bigl\}.
  \label{eq:CenterParam}
\end{equation}
The geometrical interpretation of this parameterization of the
center manifold is illustrated in \fig{fig:AmpMap}.

The flow in $\Wc$ defined by \eqn{eq:Ode} is determined
through \eqn{eq:CenterParam} by solutions to the differential
equation in $\vec{z} \in \Ec$
\begin{align}
  \Bigl(
    \vec{I} + \difp{\vec{h}}{\vec{z}}
  \Bigr) 
  \mdot \dotb{z} =
  \vec{J} \mdot \vec{z} + \vec{J} \mdot \vec{h}(\vec{z}) + 
  \vec{f}\bigl(\vec{z} + \vec{h}(\vec{z})\bigr),
  \label{eq:AmplInter}
\end{align}
in which $\vec{I}$ is the unit tensor. \Eqn{eq:AmplInter} may
be transformed to the differential equation
\begin{align}
  \dotb{z} &= 
  \Bigl(
    \vec{I} + \difp{\vec{h}}{\vec{z}}
  \Bigr)^{-1}\!\! \mdot 
  \Bigl(
    \vec{J} \mdot \vec{z} + \vec{J} \mdot \vec{h}(\vec{z}) + 
    \vec{f}\bigl(\vec{z} + \vec{h}(\vec{z})\bigr)
  \Bigr)
  \label{eq:AmplNorm}
\end{align}
which we formally express as
\begin{equation}
  \dotb{z} = 
  \vec{J} \mdot \vec{z} + \vec{g}(\vec{z}),
  \label{eq:AmplFormal}
\end{equation}
with the nonlinear part $\vec{g}(\vec{z})$ still to be
determined.

We now represent the vector $\vec{z} \in \Ec$ by coordinates
$z_1,\dots,z_r$ in the basis of critical eigenvectors
\begin{equation}
  \vec{z} = \sum_{j=1}^r z_j \vec{u}_j.
  \label{eq:ZCoord}
\end{equation}
We can then Taylor expand the relevant functions defined on
$\Ec$ in the coordinates $z_j$ as follows
\begin{subequations}
  \begin{align}
    \label{eq:FExpansion}%
    \vec{f}\bigl(\vec{z} + \vec{h}(\vec{z})\bigr) &=
    \sum_\vec{p} \vec{f}_\vec{p} \vec{z}^\vec{p},\\
    \label{eq:HExpansion}%
    \vec{h}(\vec{z}) &=
    \sum_\vec{p} \vec{h}_\vec{p} \vec{z}^\vec{p},\\
    \vec{g}(\vec{z}) &=
    \sum_\vec{p} \vec{g}_\vec{p} \vec{z}^\vec{p}.
  \end{align}
  \label{eq:Expansions}%
\end{subequations}
Here
\begin{subequations}
  \begin{align}
    \vec{p} &= (p_1,\dots,p_r) \in \N^r_0,\\
    \vec{z}^\vec{p} &= \prod_{j=1}^r z_j^{p_j},
  \end{align}
\end{subequations}
and $\abs{\vec{p}} = \sum_{j=1}^r p_j \geq 2$ for all terms of
the nonlinear functions defined in \eqn{eq:Expansions}.  The
coefficients $\vec{f}_\vec{p}$, $\vec{h}_\vec{p}$, and
$\vec{g}_\vec{p}$ are vectors in $\R^n$ ($\C^n$ if any of the
critical eigenvalues are complex).  In terms of the particular
coordinates $z_j$, the set of differential equations from
\eqn{eq:AmplFormal} for the flow on the center manifold $\Wc$
becomes
\begin{equation}
  \dot{z}_j = \lambda_j z_j + \sum_\vec{p} (\vec{u}_j^* \mdot
  \vec{g}_\vec{p}) \vec{z}^\vec{p},
  \label{eq:AmplEqn}
\end{equation}
which we shall refer to as the \emph{amplitude equation} for
\eqn{eq:Ode} on the center manifold when the transformation
$\vec{h}(\vec{z})$ is chosen so that \eqn{eq:AmplEqn} contains
as few non-vanishing coefficients $\vec{u}_j^* \mdot
\vec{g}_\vec{p}$ as possible.  As described below, this
procedure also defines the transformation $\vec{h}(\vec{z})$.

The transformation $\vec{h}(\vec{z})$ is determined through
its coefficients $\vec{h}_\vec{p}$ in \eqn{eq:HExpansion}
which must be obtained iteratively together with the
coefficients $\vec{f}_\vec{p}$ and $\vec{g}_\vec{p}$.  This is
possible because the center subspace $\Ec$ is tangent to the
center manifold $\Wc$ at $\vec{x} = \vec{0}$.  We therefore
substitute \eqn{eq:AmplFormal} and \eqn{eq:Expansions} in
\eqn{eq:AmplInter} to get
\begin{equation}
  \begin{split}
    &
    \bigl(
      \vec{I} + 
      \sum_\vec{p} \sum_{j=1}^r \vec{h}_\vec{p} \vec{z}^\vec{p}
      \frac{p_j}{z_j} \vec{u}_j^*
    \bigr)
    \mdot 
    \bigl(
      \sum_{k=1}^r z_k \lambda_k \vec{u}_k +
      \sum_\vec{p} \vec{g}_\vec{p} \vec{z}^\vec{p}
    \bigr)
    = \\
    &
    \qquad
    \sum_{k=1}^r z_k \lambda_k \vec{u}_k +
    \vec{J} \mdot \sum_\vec{p} \vec{h}_\vec{p} \vec{z}^\vec{p} +
    \sum_\vec{p} \vec{f}_\vec{p} \vec{z}^\vec{p}.
  \end{split}
  \label{eq:AmplBase}
\end{equation}
The equation for the linear terms has already been anticipated
in \eqn{eq:AmplFormal}.  For the nonlinear terms,
\eqn{eq:AmplBase} must be satisfied for each term indicated by
$\vec{p}$ separately.  We shall introduce a notation for the
coefficient to $\vec{z}^\vec{p}$ from the product of the two
sums over $\vec{p}$ on the left taken together with the last
term on the right-hand side of \eqn{eq:AmplBase}, namely
\begin{equation}
  \BPhi_\vec{p} = \vec{f}_\vec{p} -
  \sum_{\vec{p}'} \vec{h}_{\vec{p}'} 
  \sum_{j=1}^r p_j'
  \vec{u}_j^* \mdot
  \vec{g}_{(\vec{p}-\vec{p}'+\bdl_{\!j})}.
  \label{eq:PhiDef}
\end{equation}
Here the sum over $\vec{p}'$ is taken over all ordered sets
$\vec{p}' = (p_1',\dots,p_r')$ for which the terms are
defined, whereas the set $\bdl_{\!j}$ has $\delta_{jk}$ as its
$k$'th component.

From \eqn{eq:AmplBase}, we then get the coefficient
$\vec{g}_\vec{p}$ as
\begin{equation}
  \vec{g}_\vec{p} = 
  \Bigl(
    \vec{J}  - \sum_{j=1}^r p_j \lambda_j \vec{I}
  \Bigr) \mdot 
  \vec{h}_\vec{p} + \BPhi_\vec{p}.
  \label{eq:GSolve}
\end{equation}
Thus, the coefficient $\vec{g}_\vec{p}$ can be eliminated if
the coefficient $\vec{h}_\vec{p}$ of the transformation
$\vec{h}(\vec{z})$ can be chosen in such a way that the
right-hand side of \eqn{eq:GSolve} is zero, that is if
\begin{equation}
  \Bigl(
  \vec{J} - \sum_{j=1}^r p_j \lambda_j \vec{I}
  \Bigr) \mdot 
  \vec{h}_\vec{p} = -\BPhi_\vec{p},
  \label{eq:LinearEq}
\end{equation}
is solvable for $\vec{h}_\vec{p}$.  By biorthogonality,
\eqn{eq:LinearEq} splits into two independent equations for
components in $\Ec$ and components in the generalized
eigenspace complement to $\Ec$.  In the complement subspace,
\eqn{eq:LinearEq} can always be satisfied.  For components in
$\Ec$, \eqn{eq:LinearEq} can be solved for any component
$\vec{u}_i^* \mdot \vec{h}_\vec{p}$ of $\vec{h}_\vec{p}$ in
the eigenvector basis $\vec{u}_i$ for which $\sum_{j=1}^r p_j
\lambda_j \neq \lambda_i$.  For each of these components, we
may eliminate the corresponding term $\vec{u}_i^* \mdot
\vec{g}_\vec{p}$ of \eqn{eq:AmplEqn} by choosing $\vec{u}_i^*
\mdot \vec{h}_\vec{p}$ as the solution to \eqn{eq:LinearEq},
namely
\begin{equation}
  \vec{u}_i^* \mdot \vec{h}_\vec{p} = 
  \frac{-\vec{u}_i^* \mdot \BPhi_\vec{p}}
  {\lambda_i - \sum_{j=1}^r p_j \lambda_j}.
\end{equation}

In contrast, a component $\vec{u}_i^* \mdot \vec{g}_\vec{p}$
of $\vec{g}_\vec{p}$ cannot be eliminated by any choice of
transformation $\vec{h}(\vec{z})$ if the resonance condition
\begin{equation}
  {\sum_{j=1}^r p_j \lambda_j} = \lambda_i
  \label{eq:ResCond}
\end{equation}
applies.  The condition \eqn{eq:ResCond} is identical to the
one appearing in a somewhat different setting in the
traditional derivation of the normal form
theorem~\cite{Arn83,ArrPla90}, and the essential contents are
of course the same in the two descriptions.  The resonant
terms must appear in the amplitude equation \eqn{eq:AmplEqn}
and are determined by \eqn{eq:GSolve} once the corresponding
components $\vec{u}_i^* \mdot \vec{h}_\vec{p}$ of the
transformation $\vec{h}(\vec{z})$ have been chosen.  We set
each of them equal to zero so any non-vanishing (resonant)
coefficient becomes $\bigl($from \eqn{eq:GSolve}$\bigl)$
\begin{equation}
  \vec{u}_i^* \mdot \vec{g}_\vec{p} =
  \vec{u}_i^* \mdot \BPhi_\vec{p}.
  \label{eq:AmplCoeff}
\end{equation}
The amplitude equation can now be written more explicitly as
\begin{equation}
  \dot{z}_i = \lambda_i z_i + 
  \sum_\vec{p}
  (\vec{u}_i^* \mdot \BPhi_\vec{p}) \vec{z}^\vec{p},
  \quad i=1,\dots,p,
  \label{eq:AmplEqnFinal}
\end{equation}
in which the sum for the $i$'th amplitude $z_i$ is taken over
all sets $\vec{p}$ that satisfy the resonance condition
\eqn{eq:ResCond}.

In practice, calculation of the amplitude equation
\eqn{eq:AmplEqnFinal} does not require a decomposition of
\eqn{eq:LinearEq} as we did above.  For any given $\vec{p}$,
we may solve \eqn{eq:LinearEq} directly for all non-resonant
components of $\vec{h}_\vec{p}$ in terms of the part of
$\BPhi_\vec{p}$ biorthogonal to the null space of $(\vec{J} -
\sum_{j=1}^r p_j \lambda_j \vec{I})$, so $\vec{h}_\vec{p}$ is
completely determined by the set of equations
\begin{subequations}
  \label{eq:LinearEqRes}
  \begin{align}
    \label{eq:LinearEqResA}
    \Bigl(
    \vec{J}  - \sum_{j=1}^r p_j \lambda_j \vec{I}
    \Bigr) \mdot  
    \vec{h}_\vec{p} &= 
    - \vec{Q}_\vec{p} \mdot \BPhi_\vec{p},\\
    \label{eq:LinearEqResB}
    \vec{R}_\vec{p} \mdot 
    \vec{h}_\vec{p} &= \vec{0}.
  \end{align}
\end{subequations}
Here the projections $\vec{R}_\vec{p}$ and $\vec{Q}_\vec{p}$
are defined for any $\vec{x} \in \R^n$ as
\begin{equation}
  \label{eq:ProjDef}
  \vec{R}_\vec{p} \mdot \vec{x} = 
  (\vec{I} - \vec{Q}_\vec{p}) \mdot \vec{x} =
  \sum_i
  (\vec{u}_i^* \mdot \vec{x}) \vec{u}_i,
\end{equation}
where the sum is taken over all integers $i$ for which the
resonance condition \eqn{eq:ResCond} is satisfied at given
order $\vec{p}$.  In practice, the solution of
\eqn{eq:LinearEqRes} is conveniently obtained by means of
singular value decomposition \cite{GolLoan96}.

We shall refer to the transformation $\vec{x} = \vec{z} +
\vec{h}(\vec{z})$ from $\Ec$ to $\Wc$ with $\vec{h}(\vec{z})$
determined by \eqns{eq:HExpansion}{eq:LinearEqRes} as the
\emph{amplitude transformation} associated with the
\eqn{eq:Ode}.

The problem of finding the non-vanishing coefficients of the
amplitude equation \eqn{eq:AmplEqnFinal} and the coefficients
$\vec{h}_\vec{p}$ of the amplitude transformation leading to
it, is solved order by order for increasing $\abs{\vec{p}} =
\sum_{j=1}^r p_j $.  For any given $\vec{p}$ with
$\abs{\vec{p}} \geq 2$, the non-vanishing components of the
vector coefficient $\vec{g}_\vec{p}$ of the amplitude equation
are determined by the \emph{resonant part} of $\BPhi_\vec{p}$
whereas $\vec{h}_\vec{p}$ is obtained as the solution to
\eqn{eq:LinearEqRes} in terms of the \emph{non-resonant part}
of $\BPhi_\vec{p}$.  We note that, for any $\vec{p},i$, either
$\vec{u}_i^* \mdot \vec{g}_\vec{p}$ or $\vec{u}_i^* \mdot
\vec{h}_\vec{p}$ must vanish.

The iterative solution is possible because $\BPhi_\vec{p}$ is
determined by lower order coefficients of $\vec{g}_\vec{p}$
and $\vec{h}_\vec{p}$.  This fact is obvious since the sum
over $\vec{p}'$ in \eqn{eq:PhiDef} only involves coefficients
of orders within the range $2 \leq \vec{p}' \leq
\abs{\vec{p}}-1$, and since the Taylor expansion of
$\vec{f}(\vec{x})$ contains no linear terms.  Consequently,
the first term, $\vec{f}_\vec{p}$, on the right-hand side of
\eqn{eq:PhiDef} only contains components $\vec{h}_{\vec{p}'}$
of orders $2,\dots,\abs{\vec{p}}-1$ through
\eqn{eq:FExpansion}.

The amplitude equation \eqn{eq:AmplEqnFinal} and the linear
system of equations \eqn{eq:LinearEqRes} together with
expression \eqn{eq:PhiDef} for $\BPhi_\vec{p}$, constitute the
main result of this section.  Their practical use for a number
of specific bifurcations will be demonstrated in
\sect{sec:GenExamp}.  First we need to extend the theory to
systems defined at a finite distance from the bifurcation
point.  This we do next.

\section{Unfoldings of amplitude equations}
\label{sec:Unfold}
For physical systems, one is often interested in studying the
behavior of a dynamical system in a vicinity of a certain
bifurcation point.  We therefore consider a dynamical system
\begin{equation}
  \dotb{x} = \vec{F}(\vec{x},\bmu) =
  \vec{J} \mdot \vec{x} +
  \vec{f}(\vec{x},\bmu),
  \label{eq:OdePar}
\end{equation}
depending on a set of parameters $\bmu \in \R^s$, in such a
way that it contains the system \eqn{eq:Ode} considered
previously as the special case $\bmu = \vec{0}$.  In physical
problems, the motion in a ``slow manifold'' at $\bmu \neq
\vec{0}$ smoothly developed from the center manifold at $\bmu
= \vec{0}$, is of special importance.  There are several ways
of approaching this problem.  One may treat the motion in an
unstable manifold in the same way as described in the
(previous) \sect{sec:Flows} for the center manifold with but a
minor modification.  We shall comment on this idea in
\sect{sec:UnInt}.

Here we use an approach described in \cite{GuckHolmes} and
extend the differential equation~(\ref{eq:OdePar}) as
\begin{equation}
  \begin{split}
    \dotb{x} &= 
    \vec{J} \mdot \vec{x} +
    \vec{f}(\vec{x},\bmu),\\
    \dot{\bmu} &= \vec{0},
  \end{split}
  \label{eq:OdeParTwo}
\end{equation}
in which $(\vec{x},\bmu) \in \R^{n+s}$.  Note that we treat
the term $\vec{J} \mdot \vec{x}$ linear in $\vec{x}$
explicitly in \eqn{eq:OdeParTwo} whereas all other terms are
contained in $\vec{f}(\vec{x},\bmu)$.

We are interested in the motion in the center manifold of the
extended system \eqn{eq:OdeParTwo}, which represents the slow
motion for finite $\bmu$ near $\bmu = \vec{0}$, desired for an
efficient description of the physical problem.  Unfortunately,
we cannot use the results of \sect{sec:Flows} directly because
the linear part of the vector field in \eqn{eq:OdeParTwo} need
not have $r+s$ independent critical eigenvectors.
Nevertheless, the structure of \eqn{eq:OdeParTwo} is so simple
that we can use essentially the same approach as in
\sect{sec:Flows}.  We sketch the arguments and introduce the
necessary notation in order to analyze the extended system.

We consider only a dependence of the vector field on the
parameter $\bmu$ for which $(\vec{0},\bmu) \in \R^{n+s}$
belongs to the center subspace of \eqn{eq:OdeParTwo} at
$(\vec{x},\bmu) = (\vec{0},\vec{0})$.  Then the center
manifold of \eqn{eq:OdeParTwo} may be parameterized by points
$(\vec{z},\bmu) \in \Ec \times \R^s$ in the center subspace of
the extended system through a transformation $\vec{h}: \R^n
\times \R^s \rightarrow \R^n$:
\begin{equation}
  \Wc = 
  \bigl\{ 
  (\vec{x},\bmu) \;\bigl|\;
  \vec{x} = \vec{z} + \vec{h}(\vec{z},\bmu)
  \bigl\}.
  \label{eq:CenterParamUn}
\end{equation}
This means that the flow in $\Wc$ defined by
\eqn{eq:OdeParTwo} is determined by the differential equation
\begin{equation}
  \begin{split}
    \dotb{z} &= 
    \Bigl(
      \vec{I} + \difp{\vec{h}}{\vec{z}}
    \Bigr)^{-1}\!\! \mdot 
    \Bigl(
      \vec{J} \mdot \vec{z} +
      \vec{J} \mdot \vec{h}(\vec{z},\bmu)) +
      \vec{f}\bigl(\vec{z} + \vec{h}(\vec{z},\bmu), \bmu\bigr)
    \Bigr),\\
    \dot{\bmu} &= \vec{0}.
  \end{split}
  \label{eq:AmplNormUn}
\end{equation}
Clearly, this differential equation can be written in the form
\begin{equation}
  \dotb{z} = 
  \vec{J} \mdot \vec{z} + \vec{g}(\vec{z},\bmu).
  \label{eq:AmplFormalUn}
\end{equation}
We express $\vec{z}$ in a basis of eigenvectors of $\vec{J}$
as $\vec{z} = \sum_{j=1}^r z_j \vec{u}_j$, and expand the
functions $\vec{f}$, $\vec{h}$, and $\vec{g}$ in a manner
similar to \eqn{eq:Expansions} as
\begin{subequations}
  \begin{align}
    \vec{f}\bigl(\vec{z} + \vec{h}(\vec{z},\bmu),\bmu\bigr) &=
    \sum_\vec{pq} \vec{f}_\vec{pq} \vec{z}^\vec{p}\bmu^\vec{q},\\
    \vec{h}(\vec{z},\bmu) &=
    \sum_\vec{pq} \vec{h}_\vec{pq} \vec{z}^\vec{p}\bmu^\vec{q},\\
    \vec{g}(\vec{z},\bmu) &=
    \sum_\vec{pq} \vec{g}_\vec{pq} \vec{z}^\vec{p}\bmu^\vec{q},
  \end{align}
  \label{eq:ExpUnfold}%
\end{subequations}
in which 
\begin{subequations}
  \begin{align}
    \vec{p} &= (p_1,\dots,p_r) \in \N^r_0, \quad
    \vec{q} = (q_1,\dots,q_s) \in \N^s_0,\\
    \vec{z}^\vec{p} &= \prod_{j=1}^r z_j^{p_j}, \quad
       \bmu^\vec{q} = \prod_{k=1}^s \mu_k^{q_k}.
  \end{align}
\end{subequations}
Note that we have introduced the short-hand notation
$\vec{pq}$ for the index $(\vec{p},\vec{q})$.  Here,
$\abs{\vec{p}} + \abs{\vec{q}} \geq 1$, with the cases
$\abs{\vec{p}} = 1$, $\abs{\vec{q}} = 0$ excluded.  Since
\eqnto{eq:CenterParamUn}{eq:ExpUnfold} each has the same
structure as in section~\ref{sec:Flows}, we arrive at the
following equation for the vector $\vec{g}_\vec{pq}$ which
must be satisfied for each separate order~$(\vec{p},\vec{q})$
\begin{equation}
  \vec{g}_\vec{pq} = 
  \Bigl(
    \vec{J}  - \sum_{j=1}^r p_j \lambda_j \vec{I}
  \Bigr) \mdot 
  \vec{h}_\vec{pq} +
  \BPhi_\vec{pq}.
  \label{eq:GSolveUnfold}
\end{equation}
The vector $\BPhi_\vec{pq} \in \R^n$ is defined by a
straightforward extension of \eqn{eq:PhiDef}, namely
\begin{equation}
  \BPhi_\vec{pq} = \vec{f}_\vec{pq} -
  \sum_{\vec{p}'\vec{q}'} 
  \vec{h}_{\vec{p}'\vec{q}'} 
  \sum_{j=1}^r p_j'
  \vec{u}_j^* \mdot
  \vec{g}_{(\vec{p}-\vec{p}'+\bdl_{\!j})(\vec{q}-\vec{q}')}.
  \label{eq:PhiDefUnfold}
\end{equation}
Consequently, the vector $\vec{g}_\vec{pq}$ will vanish
provided that $\vec{h}_\vec{pq}$ is chosen as a solution to
the equation
\begin{subequations}
  \label{eq:LinearEqUn}
  \begin{align}
    \label{eq:LinearEqUnA}
    \Bigl(
      \vec{J}  - \sum_{j=1}^r p_j \lambda_j \vec{I}
    \Bigr) \mdot 
    \vec{h}_\vec{pq} &= 
    -\vec{Q}_\vec{p} \mdot \BPhi_\vec{pq},\\
    \vec{R}_\vec{p} \mdot \vec{h}_\vec{pq} &= \vec{0}.
    \label{eq:LinearEqUnB}
  \end{align}
\end{subequations}
Observe now that \eqn{eq:LinearEqUn} has exactly the same form
as \eqn{eq:LinearEqRes}.  We may therefore conclude that the
unfolding of the amplitude equation in \eqn{eq:AmplEqnFinal}
simply becomes
\begin{equation}
  \dot{z}_i = \lambda_i z_i + 
  \sum_{\vec{pq}}
  (\vec{u}_i^* \mdot \BPhi_\vec{pq}) \vec{z}^\vec{p}\bmu^\vec{q},
  \quad i=1,\dots,p,
  \label{eq:AmplUnfold}
\end{equation}
where the sum for the $i$'th component is taken over all sets
$(\vec{p},\vec{q})$ where resonance occurs.  For $\bmu =
\vec{0}$, only terms with $\abs{\vec{q}} = 0$ contribute, and
we recover the results of the previous section.  For general
$\bmu$, \eqn{eq:AmplUnfold} is an extension of
\eqn{eq:AmplEqnFinal} which incorporates the parametric
dependence on $\bmu$ of the resonant terms.

\Eqns{eq:GSolveUnfold}{eq:PhiDefUnfold} thus constitute a
recurrence relation for determining $\vec{g}_{pq}$ and
$\vec{h}_{pq}$.  One must first determine the coefficients for
$\abs{\vec{q}} = 0$ for all orders $\vec{p}$ up to $p_0 + 1$
where $p_0$ is the maximum value of $\abs{\vec{p}}$ to be
included for any $\vec{q}$.  Subsequently, the coefficients
are determined for all $q$ with $\abs{\vec{q}} = 1$ in the
same way.

On the center manifold, the functions $\vec{g}(\vec{z},\bmu)$
and $\vec{h}(\vec{z},\bmu)$ are at least quadratic in
$\vec{z}$.  However, the unfoldings may contain nonvanishing
coefficients $\vec{g}_{pq}$ and $\vec{h}_{pq}$ with
$\abs{\vec{p}} = 0,1$ when $\abs{\vec{q}} \geq 1$.  In other
words, the functions $\vec{g}(\vec{z},\bmu)$ and
$\vec{h}(\vec{z},\bmu)$ may contain terms depending on $\bmu$
which are constant or linear in $\vec{z}$ which now briefly
discuss.  For $\abs{\vec{p}} = 0$ and $\abs{\vec{q}} = 1$, we
have
\begin{equation}
  \vec{J} \mdot \vec{h}_{\vec{0}\bdl_{\!k}} =
  -\vec{Q}_\vec{0} \mdot
  \BPhi_{\vec{0}\bdl_{\!k}}, \quad \text{for \ } k = 1,\dots,q.
  \label{eq:UnfoldZero}
\end{equation}
with $\bdl_{\!k}$ defined below \eqn{eq:PhiDef}.  If
$\lambda_i \neq 0$, the corresponding coefficient $\vec{u}_i^*
\mdot \vec{g}_{\vec{0}\bdl_{\!k}}$ will vanish from the
amplitude equation provided that the coefficient $\vec{u}_i^*
\mdot \vec{h}_{\vec{0}\bdl_{\!k}}$ is chosen as
\begin{equation}
  \vec{u}_i^* \mdot \vec{h}_{\vec{0}\bdl_{\!k}} = 
  -\frac{\vec{u}_i^* \mdot \BPhi_{\vec{0}\bdl_{\!k}}}{\lambda_i}.
\end{equation}
On the other hand, if $\lambda_i = 0$ we choose $\vec{u}_i^*
\mdot \vec{h}_{\vec{0}\bdl_{\!k}} = 0$ by
\eqn{eq:LinearEqUnB}, and get the following contribution from
$\vec{u}_i^* \mdot \vec{g}_{\vec{0}\bdl_{\!k}}$ in the
amplitude equation
\begin{equation}
  \label{eq:UnfoldZeroG}
  \vec{u}_i^* \mdot \vec{g}_{\vec{0}\bdl_{\!k}} =
  \vec{u}_i^* \mdot \BPhi_{\vec{0}\bdl_{\!k}}.
\end{equation}
At the order $\abs{\vec{p}} = 1$, $\abs{\vec{q}} = 1$, we find
that the following equation must hold for each ordered
pair~$(\bdl_{\!j},\bdl_{\!k})$
\begin{equation}
  (\vec{J}  - \lambda_j \vec{I}) \mdot 
  \vec{h}_{\bdl_{\!j} \! \bdl_{\!k}} =
  -\vec{Q}_{\bdl_{\!j}} \!\mdot 
  \BPhi_{\bdl_{\!j} \! \bdl_{\!k}}.
  \label{eq:UnfoldOne}
\end{equation}
For $i=j$, we always have $\vec{u}_j^* \mdot\, (\vec{J} -
\lambda_j \vec{I}) \mdot \vec{h}_{\bdl_{\!j} \! \bdl_{\!k}} =
0$, implying that we get a contribution to the amplitude
equation given by \mbox{$\vec{u}_j^* \mdot
\vec{g}_{\bdl_{\!j}\bdl_{\!k}} = \vec{u}_j^* \mdot
\BPhi_{\bdl_{\!j}\bdl_{\!k}}$} and \mbox{$\vec{u}_j^* \mdot
\vec{h}_{\bdl_{\!j}\bdl_{\!k}} = 0$} from
\eqn{eq:LinearEqUnB}.  In the asymmetric case $i \neq j$ we
have
\begin{subequations}
  \begin{align}
    \lambda_i &\neq \lambda_j: &
    \vec{u}_i^* \mdot \vec{h}_{\bdl_{\!j} \! \bdl_{\!k}} &= 
    -\frac{\vec{u}_i^* \mdot \BPhi_{\bdl_{\!j} \! \bdl_{\!k}}}
    {\lambda_i-\lambda_j}, &
    \vec{u}_i^* \mdot \vec{g}_{\bdl_{\!j} \! \bdl_{\!k}} &= 0,
    \\
    \lambda_i &   = \lambda_j: &  
    \vec{u}_i^* \mdot \vec{h}_{\bdl_{\!j} \! \bdl_{\!k}} &= 0, &
    \vec{u}_i^* \mdot \vec{g}_{\bdl_{\!j} \! \bdl_{\!k}} &= 
    \vec{u}_i^* \mdot \BPhi_{\bdl_{\!j} \! \bdl_{\!k}}.
  \end{align}
\end{subequations}

We shall illustrate the use of these results in
\sect{sec:GenExamp} and explain their meaning in
\sect{sec:UnInt}.

\section{Amplitude equations for iterated maps}
\label{sec:Disc}
The description of the dynamics on center manifolds for
iterated maps is very similar to that of flows presented in
\sect{sec:Flows}.  Therefore we merely give a short outline of
the theory.

A discrete system in $\R^n$ is described by a map of the form
\begin{equation}
  \vec{x} \mapsto \vec{F}(\vec{x}) = 
  \vec{J} \mdot \vec{x} + \vec{f}(\vec{x}).
  \label{eq:DiscDiff}
\end{equation}
We assume that $\vec{x} = \vec{0}$ is a non-hyperbolic fixed
point of \eqn{eq:DiscDiff} where the linear part $\vec{J}$ has
$r$ eigenvalues $\lambda_j$ with $\abs{\lambda_j} = 1$ ($j =
1,\dots,r$).  The corresponding right and left eigenvectors
are denoted by $\vec{u}_j$ and $\vec{u}_j^*$ and are assumed
to be normalized in accordance with the biorthonormal
relations in~\eqn{eq:BiOrtho}.  The $r$-dimensional center
manifold $\Wc$ of the fixed point $\vec{x} = \vec{0}$ may be
parameterized by points in the center subspace as
\begin{equation}
  \vec{x} = \vec{z} + \vec{h}(\vec{z}) \equiv \vec{K}(\vec{z}),
  \label{eq:CenterDisPar}
\end{equation}
where $\vec{x} \in \Wc$ and $\vec{z} \in \Ec$.  Inserting
\eqn{eq:CenterDisPar} into \eqn{eq:DiscDiff} yields
\begin{equation}
  \vec{z} + \vec{h}(\vec{z}) \mapsto
  \vec{J} \mdot \vec{z} + 
  \vec{J} \mdot \vec{h}(\vec{z}) +
  \vec{f}\bigl(\vec{z} + \vec{h}(\vec{z})\bigr),
  \label{eq:DiscInter}
\end{equation}
which shows that the discrete dynamics on the center manifold
is induced by the map
\begin{equation}
  \vec{z} \mapsto
  \vec{K}^{-1}
  \bigl( 
    \vec{J} \mdot \vec{z} + 
    \vec{J} \mdot \vec{h}(\vec{z}) +
    \vec{f}\bigl(\vec{z} + \vec{h}(\vec{z})\bigr)
  \bigr),
  \label{eq:DiscGeneral}
\end{equation}
in the center subspace. 

To find the map in $\Ec$ explicitly, we formally express it as
\begin{equation}
  \vec{z} \mapsto \vec{J} \mdot \vec{z} + \vec{g}(\vec{z}).
  \label{eq:DiscFormal}
\end{equation}
As in the continuous case, we express $\vec{z}$ in terms of
coordinates based on the right eigenvectors $\vec{u}_j$,
\emph{viz.}\ $\vec{z} = \sum_{j=1}^r z_j \vec{u}_j$.  The
functions $\vec{f}\bigl(\vec{z} + \vec{h}(\vec{z})\bigr)$,
$\vec{h}(\vec{z})$, and $\vec{g}(\vec{z})$ can then be Taylor
expanded as
\begin{subequations}
  \begin{align}
    \vec{f}\bigl(\vec{z} + \vec{h}(\vec{z})\bigr) &=
    \sum_\vec{p} \vec{f}_\vec{p} \vec{z}^\vec{p},\\
    \label{eq:DiscExpH}%
    \vec{h}(\vec{z}) &=
    \sum_\vec{p} \vec{h}_\vec{p} \vec{z}^\vec{p},\\
    \vec{g}(\vec{z}) &=
    \sum_\vec{p} \vec{g}_\vec{p} \vec{z}^\vec{p},
  \end{align}
  \label{eq:DiscExpansions}%
\end{subequations}
and the map \eqn{eq:DiscFormal} becomes
\begin{equation}
  z_i \mapsto \lambda_i z_i + 
  \sum_{\vec{p}} (\vec{u}_i^* \mdot 
  \vec{g}_\vec{p}) \vec{z}^\vec{p}.
  \label{eq:AmplEqnDisc}
\end{equation}
When the transformation $\vec{h}(\vec{z})$ is chosen so that
\eqn{eq:AmplEqnDisc} contains as few non-vanishing
coefficients $\vec{u}_j^* \mdot \vec{g}_\vec{p}$ as possible,
we shall refer to this equation as the amplitude equation for
the map in \eqn{eq:DiscDiff}.

To determine $\vec{h}(\vec{z})$, we first substitute
\eqn{eq:DiscFormal} for $\vec{z}$ on the left-hand side of
\eqn{eq:DiscInter}.  Since the resulting expression is equal
to the right-hand side of \eqn{eq:DiscInter}, we obtain
\begin{equation}
  \vec{J} \mdot \vec{z} + \vec{g}(\vec{z}) +
  \vec{h}\bigl(
    \vec{J} \mdot \vec{z} + \vec{g}(\vec{z})
  \bigr) =
  \vec{J} \mdot \vec{z} + 
  \vec{J} \mdot \vec{h}(\vec{z}) +
  \vec{f}\bigl(\vec{z} + \vec{h}(\vec{z})\bigr).
  \label{eq:DiscInter2}
\end{equation}
Here we note that the term $ \vec{h}\bigl( \vec{J} \mdot
\vec{z} + \vec{g}(\vec{z}) \bigr)$ may be written as
\begin{align}
  \vec{h}\bigl(
    \vec{J} \mdot \vec{z} + \vec{g}(\vec{z})
  \bigr) &=
  \sum_\vec{p} \vec{h}_\vec{p} \prod_{j=1}^r 
  \bigl(
    \lambda_j z_j + \sum_{\vec{p}'} 
    (\vec{u}_j^* \mdot \vec{g}_{\vec{p}'}) \vec{z}^{\vec{p}'}
  \bigr)^{p_j},\nonumber\\
  &= 
  \sum_\vec{p} \vec{h}_\vec{p} \vec{z}^\vec{p} 
  \prod_{j=1}^r \lambda_j^{p_j} +
  \sum_\vec{p} \BPsi_\vec{p} \vec{z}^\vec{p}.
  \label{eq:ProdExp}
\end{align}
The last step defines the vector $\BPsi_\vec{p} \in \R^n$
which abbreviates the (rather complicated) coefficient that
remains at order $\vec{p}$ when the term $\sum_\vec{p}
\vec{h}_\vec{p} \vec{z}^\vec{p} \prod_{j=1}^r \lambda_j^{p_j}$
is taken out.  Substitution of \mbox{$\vec{z} = \sum_{j=1}^r
z_j \vec{u}_j$} together with \eqn{eq:DiscExpansions}, and
\eqn{eq:ProdExp} in \eqn{eq:DiscInter2} followed by a simple
rearrangement of terms then yields the following expression
for the coefficient vector~$\vec{g}_\vec{p}$
\begin{equation}
  \vec{g}_\vec{p} =
  \Bigl(
    \vec{J} - \prod_{j=1}^r\lambda_j^{p_j}\vec{I}
  \Bigr) \mdot \vec{h}_\vec{p} +
  \BPhi_\vec{p},
  \label{eq:DiscPhiDef}
\end{equation}
where $\BPhi_\vec{p} = \vec{f}_\vec{p} - \BPsi_\vec{p}$.
Consequently, the coefficient vector $\vec{g}_\vec{p}$ can be
eliminated if $\vec{h}_\vec{p}$ is chosen as a solution to the
equation
\begin{equation}
  \bigl(
    \vec{J} - \prod_{j=1}^r\lambda_j^{p_j}\vec{I}
  \bigr) \mdot \vec{h}_\vec{p} =
  -\BPhi_\vec{p}.
  \label{eq:DiscLinearEq}
\end{equation}
However, a component $\vec{u}_i^* \mdot \vec{g}_\vec{p}$ of
$\vec{g}_\vec{p}$ cannot be removed by any choice of
$\vec{h}_\vec{p}$ if the resonance condition
\begin{equation}
  \prod_{j=1}^r \lambda_j^{p_j} = \lambda_i,
  \label{eq:DiscResonance}
\end{equation}
is satisfied. 

Using a series of arguments analogous to the derivation of the
amplitude equation for flows given in section~\ref{sec:Flows},
one easily obtains the following recurrence relation for the
coefficients $\vec{g}_\vec{p}$ and $\vec{h}_\vec{p}$ of the
amplitude equation and the associated transformation
\begin{equation}
  \vec{g}_\vec{p} =
  \Bigl(
    \vec{J} - \prod_{j=1}^r\lambda_j^{p_j}\vec{I}
  \Bigr) \mdot \vec{h}_\vec{p} +
  \BPhi_\vec{p}.
\end{equation}
We may therefore conclude that the unfolding of the amplitude
equation \eqn{eq:AmplEqnDisc} becomes
\begin{equation}
  z_i \mapsto \lambda_i z_i + 
  \sum_{\vec{p}} (\vec{u}_i^* \mdot \BPhi_\vec{p}) \vec{z}^\vec{p},
  \quad i=1,\dots,p,
  \label{eq:DiscAmplEqnFinal}
\end{equation}
where the sum is taken over all resonant sets $\vec{p}$ for
the $i$'th component.  Furthermore, all resonant components
$\vec{u}_i^* \mdot \vec{h}_\vec{p}$ of $\vec{h}_\vec{p}$ are
set equal to zero whereas non-resonant terms are removed by
choosing the corresponding coefficient $\vec{u}_i \mdot
\vec{h}_\vec{p}$ of $\vec{h}_\vec{p}$ as
\begin{equation}
  \vec{u}_i^* \mdot \vec{h}_\vec{p} = 
  \frac{-\vec{u}_i^* \mdot \BPhi_\vec{p}}
  {\lambda_i - \prod_{j=1}^r \lambda_j^{p_j}}.
\end{equation}
Again, we note that the resonance condition
\eqn{eq:DiscResonance} is identical to the one that appears in
the traditional normal form derivation for diffeomorphisms
\cite{Arn83,ArrPla90}.

For a map depending on a set of parameters $\bmu \in \R^s$,
\begin{equation}
  \vec{x} \mapsto \vec{F}(\vec{x},\bmu) = 
  \vec{J} \mdot \vec{x} + \vec{f}(\vec{x},\bmu),
  \label{eq:DiscDiffPar}
\end{equation}
we use the same approach as for flows and study the extended
system
\begin{equation}
  \begin{split}
    \vec{x} &\mapsto \vec{J} \mdot \vec{x} + \vec{f}(\vec{x},\bmu),\\
    \bmu    &\mapsto \bmu.
  \end{split}
  \label{eq:DiscDiffUn}
\end{equation}
Using expansions \eqn{eq:ExpUnfold} and an approach completely
analogous to the one discussed above, we find a recurrence
relation 
\begin{equation}
  \vec{g}_\vec{pq} =
  \Bigl(
    \vec{J} - \prod_{j=1}^r\lambda_j^{p_j}\vec{I}
  \Bigr) \mdot \vec{h}_\vec{pq} +
  \BPhi_\vec{pq}.
  \label{eq:DiscPhiDefUn}
\end{equation}
with $\BPhi_{pq} = \vec{f}_{pq} - \BPsi_{pq}$ where the
definition of $\BPsi_{pq}$ is analogous to that of
$\BPsi_{p}$.  An example of the calculation of $\BPhi_{pq}$
for the period doubling bifurcation in \sect{ssec:PerDouble}
illustrates the use of the formula.  In conclusion, the
unfolded amplitude equation for the iterated map
\eqn{eq:DiscDiffUn} becomes
\begin{equation}
  z_i \mapsto \lambda_i z_i + 
  \sum_\vec{pq}
  (\vec{u}_i^* \mdot \BPhi_\vec{pq}) \vec{z}^\vec{p}\bmu^\vec{q},
  \quad i=1,\dots,r,
  \label{eq:DiscAmplUnfold}
\end{equation}
where the sum is taken over all sets $(\vec{p},\vec{q})$ for
which resonance occurs for the $i$'th component.

The coefficients $\vec{h}_\vec{pq}$ of the unfolded amplitude
transformation are determined as solutions to the linear
equations
\begin{subequations}
  \begin{align}
    \bigl(
    \vec{J} - \prod_{j=1}^r\lambda_j^{p_j}\vec{I}
    \bigr) \mdot \vec{h}_\vec{pq} &=
    -\vec{Q}_\vec{p} \mdot \BPhi_\vec{pq},\\
    \vec{R}_\vec{p} \mdot \vec{h}_\vec{pq} &= \vec{0}.
  \end{align}
  \label{eq:LinearEqResMap}
\end{subequations}

As for \eqn{eq:LinearEqRes}, the linear system
\eqn{eq:LinearEqResMap} is solved in practice by application
of a singular value decomposition.

In \sect{ssec:PerDouble} below,
we demonstrate the use of these results for the period
doubling bifurcation.

\section{Amplitude equations for simple bifurcations}
\label{sec:GenExamp}
In this section, we derive explicit expressions for the
coefficients of the amplitude equation and the associated
transformation $\vec{h}(\vec{z},\bmu)$ to lowest non-trivial
order for some simple bifurcations.  For this purpose, we
represent the vector field on the right-hand side of
\eqn{eq:OdePar} and \eqn{eq:DiscDiffPar} as Taylor expansions
in some physically relevant coordinates $x_j$ of $\vec{x} \in
\R^n$ and $\mu_j$ of $\bmu \in \R^s$.  In terms of $\vec{x}$
and $\bmu$ the expansion of the vector field \eqn{eq:OdePar}
takes the form
\begin{equation}
  \vec{F}(\vec{x},\bmu) = 
  \vec{J} \mdot \vec{x} +
  \vec{F}_{\bmu} \mdot \bmu +
  \vec{F}_{\vec{x}\bmu}(\vec{x},\bmu) +
  \tfrac{1}{2!} \Dfxx  (\vec{x},\vec{x}) +
  \tfrac{1}{3!} \Dfxxx (\vec{x},\vec{x},\vec{x}) + \dotsb,
  \label{eq:NonlinExp}
\end{equation}
in concise notation with 
\begin{equation}
  \begin{split}
    \Dfxx(\vec{x},\vec{x}) &= 
    \sum_{i,j=1}^{n} 
    \frac{\partial^2 \vec{F}}{\partial x_i\partial x_j} x_ix_j,\qquad
    \Dfxxx(\vec{x},\vec{x},\vec{x}) = 
    \sum_{i,j,k=1}^{n} 
    \frac{\partial^3 \vec{F}}{\partial x_i\partial x_j\partial x_k} 
    x_ix_jx_k,\\
    \vec{F}_{\bmu} \mdot \bmu 
    &= \sum_{i=1}^{s} \frac{\partial \vec{F}}{\partial \mu_i}\mu_i,\qquad
    \vec{F}_{\vec{x}\bmu}(\vec{x},\bmu) = 
    \sum_{i=1}^{n} \sum_{j=1}^{s} \frac{\partial^2 \vec{F}}
    {\partial x_i\partial \mu_j} x_i\mu_j, \;\dots
  \end{split}
  \label{eq:CompExp}
\end{equation}
where all derivatives are evaluated at $\vec{x} = \vec{0}$ and
$\bmu = \vec{0}$.  Here, the multilinear vector functions,
$\Dfxx(\vec{x},\vec{x})$, $\Dfxxx(\vec{x},\vec{x},\vec{x})$,
\emph{etc}, are symmetric in all their arguments, so we need
not specify them any further.  We shall only consider
unfoldings linear in $\bmu$, so we need not take any terms
beyond those exhibited in \eqn{eq:NonlinExp} into account.
From part of \eqn{eq:NonlinExp} nonlinear in $\vec{x}$, namely
\begin{equation}
  \vec{f}(\vec{x},\bmu) = 
  \vec{F}_{\bmu} \mdot \bmu +
  \vec{F}_{\vec{x}\bmu}(\vec{x},\bmu) +
  \tfrac{1}{2!} \Dfxx  (\vec{x},\vec{x}) +
  \tfrac{1}{3!} \Dfxxx (\vec{x},\vec{x},\vec{x}) +
  \dotsb,
  \label{eq:NonX}
\end{equation}
we get the coefficients $\vec{f}_\vec{pq}$ by first
substituting the parameterization of the center manifold
\begin{equation}
  \vec{x} = \sum_{j=1}^r z_j \vec{u}_j + 
  \sum_\vec{pq} \vec{h}_\vec{pq} \vec{z}^\vec{p}\bmu^\vec{q}
  \label{eq:PhysRela}
\end{equation}
for $\vec{x}$ and then collecting all terms of order
$(\vec{p},\vec{q})$ (\emph{i.e.}\ terms containing
$\vec{z}^\vec{p} \bmu^\vec{q}$).  In the examples considered
here, we shall only discuss problems associated with a scalar
parametric dependence $\mu$, implying that we may simplify the
multilinear vector functions in \eqn{eq:NonlinExp} by writing
\begin{equation}
  \vec{F}_{\vec{x}^p\mu^q}
  (\vec{x},\vec{x},\dots,\vec{x},\mu,\mu,\dots,\mu) =
  \vec{F}_{\vec{x}^p\mu^q}
  (\vec{x},\vec{x},\dots,\vec{x})\mu^q.
\end{equation}

\subsection{Bifurcations at a single zero eigenvalue}
\label{ssec:RealBif}
We first look at a stationary point of the flow \eqn{eq:Ode}
where one real eigenvalue of $\vec{J}$ is zero and assume that
all other eigenvalues have non-zero real parts.  We denote the
right and left eigenvectors corresponding to $\lambda=0$ by
$\vec{u}$ and $\vec{u}^*$ respectively.  Here the center
manifold and the center subspace are both one-dimensional, and
the latter is spanned by $\vec{u}$.  There is just one
amplitude $z$, and the set of integers $\vec{p}$ becomes just
a simple index $p$.

The resonance condition \eqn{eq:ResCond} takes the form
\mbox{$\lambda=p \lambda$}, and is satisfied for all orders
\mbox{$p \geq 2$}.  So none of the nonlinear terms of the
amplitude equation vanish in general, whereas all of the
components $\vec{u}^* \mdot \vec{h}_{pq}$ do.  At the
bifurcation point, the amplitude equation \eqn{eq:AmplUnfold}
therefore has the following form
\begin{equation}
  \dot{z} = 
  g_{01} \mu + g_{11} \mu z + g_{20} z^2 +  g_{30} z^3 + \dotsb,
  \label{eq:RealFormal}
\end{equation}
in which \mbox{$g_{pq} = \vec{u}^* \mdot \vec{g}_{pq}$}.  For
the lowest order contributions to the amplitude equation and
the transformation $\vec{h}(\vec{z},\mu)$, we need first
calculate $\BPhi_{pq}$ from \eqn{eq:PhiDefUnfold}.  To obtain
the coefficients $\vec{f}_{pq}$, we substitute
\begin{equation}
  \vec{x} = \vec{u} z + \vec{h}_{01} \mu + \vec{h}_{11} \mu z 
  + \vec{h}_{20} z^2 + \vec{h}_{30} z^3 + \dotsb
\end{equation}
in the expansion \eqn{eq:NonX}
\begin{equation}
  \begin{split}
    \vec{f}(\vec{x}) &= 
    \Dfp \mu +
    \bigl( 
      \vec{F}_{\vec{x}\mu} \mdot \vec{u} + 
      \Dfxx (\vec{u},\vec{h}_{01})
    \bigr) \mu z + \\
    & \qquad
    \tfrac{1}{2}\Dfxx (\vec{u},\vec{u})z^2 +
    \bigl(
      \Dfxx (\vec{u},\vec{h}_{20}) +
     \tfrac{1}{6}\Dfxxx (\vec{u},\vec{u},\vec{u})
    \bigr) z^3 + \dotsb.
  \end{split}
\end{equation}
Thus,
\begin{subequations}
  \begin{align}
    \vec{f}_{01} &= \Dfp,\\
    \label{eq:SaddleRHSOne}
    \vec{f}_{11} &= \vec{F}_{\vec{x}\mu} \mdot \vec{u} + 
      \Dfxx (\vec{u},\vec{h}_{01}),\\
    \vec{f}_{20} &= \tfrac{1}{2}\Dfxx (\vec{u},\vec{u}),\\
    \vec{f}_{30} &= \Dfxx (\vec{u},\vec{h}_{20}) +
    \tfrac{1}{6}\Dfxxx (\vec{u},\vec{u},\vec{u}).
  \end{align}
  \label{eq:RHS}%
\end{subequations}
\textbf{Second order in $z$ ($p=2$ and $q=0$)}\\%
At second order, there is no contribution from the sum over
$p'$ and $q'$ in \eqn{eq:PhiDefUnfold}, implying that
\begin{equation}
  \BPhi_{20} = \vec{f}_{20} = \tfrac{1}{2}\Dfxx (\vec{u},\vec{u}).
  \label{eq:RealSec}
\end{equation}
We therefore get the second order coefficient of the amplitude
equation from \eqn{eq:AmplUnfold}~as
\begin{equation}
  g_{20} = \vec{u}^* \mdot \BPhi_{20} = 
  \tfrac{1}{2} \vec{u}^* \mdot \Dfxx (\vec{u},\vec{u}),
  \label{eq:RealBifG}
\end{equation}
and a linear equation for the coefficient vector
$\vec{h}_{20}$ from \eqn{eq:LinearEqUn}
\begin{subequations}
  \label{eq:RealBifTwo}
  \begin{align}
    \vec{J} \mdot \vec{h}_{20} &= -\tfrac{1}{2} \vec{Q} \mdot
    \Dfxx(\vec{u},\vec{u}),\\
    \vec{u}^* \mdot \vec{h}_{20} &= 0,
  \end{align}
\end{subequations}
with the projection $\vec{Q}$ defined through $\vec{Q} \mdot
\vec{x} = \vec{x} - (\vec{u}^* \mdot \vec{x})\vec{u}$.

\vspace{2.5\parsep}
\noindent\textbf{Third order in $z$ ($p=3$ and $q=0$)}\\
Once $\vec{g}_{20}$ and $\vec{h}_{20}$ have been obtained from
\eqn{eq:RealBifG} and \eqn{eq:RealBifTwo} respectively, we may
derive the third order contributions using
\eqn{eq:PhiDefUnfold} for $\BPhi_{30}$
\begin{align}
 \BPhi_{30} &= \vec{f}_{30} - 2 g_{20} \vec{h}_{20},\nonumber\\
         &= \Dfxx (\vec{u},\vec{h}_{20}) + 
  \tfrac{1}{6} \Dfxxx (\vec{u},\vec{u},\vec{u}) - 2 g_{20} \vec{h}_{20}.
\end{align}
We observe that $\vec{u}^* \mdot \vec{h}_{20} = 0$, and conclude
from the above expression that 
\begin{equation}
  g_{30} = \vec{u}^* \mdot \BPhi_{30} = 
  \vec{u}^* \mdot \Dfxx (\vec{u},\vec{h}_{20})
  + \tfrac{1}{6} \vec{u}^* \mdot \Dfxxx (\vec{u},\vec{u},\vec{u}),
\end{equation}
whereas the coefficient vector $\vec{h}_{30}$ is determined by
the linear equation
\begin{subequations}
  \label{eq:RealBifThree}
  \begin{align}
    \vec{J} \mdot \vec{h}_{30} &= 
    -\vec{Q} \mdot \bigl(
    \Dfxx (\vec{u},\vec{h}_{20}) +
    \tfrac{1}{6} \Dfxxx (\vec{u},\vec{u},\vec{u})
    \bigl) +
    2 g_{20} \vec{h}_{20},\\
    \vec{u}^* \mdot \vec{h}_{30} &= 0.
  \end{align}
\end{subequations}
Following this procedure, the iterative solution for $g_{p0}$ and
$\vec{h}_{p0}$ can be developed to still higher orders.

\vspace{2.5\parsep}
\noindent\textbf{Unfolding terms ($p=0$ and $q=1$)}\\
Here $\BPhi_{01} = \vec{f}_{01}$.  From \eqn{eq:RHS}, we get the
first contribution to the unfolding as
\begin{equation}
  g_{01} = \vec{u}^* \mdot \Dfp.
  \label{eq:SaddleUnOne}
\end{equation}
In order to find $\vec{h}_{01}$, we must solve the linear
equations
\begin{subequations}
  \label{eq:RealUnOne}
  \begin{align}
      \vec{J} \mdot \vec{h}_{01} &= -\vec{Q} \mdot \Dfp,\\
    \vec{u}^* \mdot \vec{h}_{01} &= 0.
  \end{align}
\end{subequations}

\vspace{2.5\parsep}
\noindent\textbf{Unfolding terms ($p=1$ and $q=1$)}\\
From \eqn{eq:PhiDefUnfold}, we find that $\BPhi_{11} =
\vec{f}_{11} - 2 g_{01} \vec{h}_{20}$.  Having obtained both
$g_{01}$ and $\vec{h}_{01}$, we may finally find the
contribution $g_{11}$ to the unfolding of the amplitude
equation \eqn{eq:RealFormal} as
\begin{equation}
  g_{11} =
  \vec{u}^* \mdot \BPhi_{11} = 
  \vec{u}^* \mdot \Dfxp \mdot \vec{u} + 
  \vec{u}^* \mdot \Dfxx (\vec{u},\vec{h}_{01}),
\end{equation}
since $\vec{u}^* \mdot \vec{h}_{20} = 0$.  The coefficient
vector $\vec{h}_{11}$ can then be obtained by solving the
linear system of equations
\begin{subequations}
  \label{eq:RealUnTwo}
  \begin{align}
    \vec{J} \mdot \vec{h}_{11} &=
    -\vec{Q} \bigl( \mdot \vec{F}_{\vec{x}\mu} \mdot \vec{u} 
    +\mdot \Dfxx (\vec{u},\vec{h}_{01}) \bigl) 
    -2 g_{01} \vec{h}_{20},\\
    \vec{u}^* \mdot \vec{h}_{11} &= 0.
  \end{align}
\end{subequations}

We now specialize the results for the amplitude equation to
three important bifurcations at a single zero eigenvalue,
including only the terms at lowest order in $z$ for $q=0,1$.
For the terms linear in $\mu$, we introduce the simpler
notation $\sigma_p = g_{p1}$, whereas we shall use $g_p =
g_{p0}$ for terms independent of $\mu$.

\subsubsection{Saddle-node bifurcation}
If $\sigma_0 \neq 0$ and $g_2 \neq 0$, a saddle-node
bifurcation occurs at $\mu = 0$, and the corresponding
amplitude equations takes the form
\begin{equation}
  \dot{z} = \sigma_0 \mu + g_2 z^2,
  \label{eq:SaddleNode}
\end{equation}
with
\begin{subequations}
  \begin{align}
    \label{eq:SaddleNonC}
    g_2 &= \tfrac{1}{2} \vec{u}^* \mdot \Dfxx
    (\vec{u},\vec{u}),\\
    \label{eq:SaddleUnC}
    \sigma_0 &= \vec{u}^* \mdot \Dfp.
  \end{align}
\end{subequations}

\subsubsection{Transcritical bifurcation}
If $\sigma_0 = 0$ but $\sigma_1 \neq 0$ and $g_2 \neq 0$, a
transcritical bifurcation occurs at $\mu = 0$, for which the
amplitude equation is
\begin{equation}
  \dot{z} = \sigma_1 \mu z + g_2 z^2,
  \label{eq:TransCrit}
\end{equation}
with $g_2$ given by \eqn{eq:SaddleNonC} and 
\begin{equation}
  \label{eq:TransUnC}
  \sigma_1 = \vec{u}^* \mdot \Dfxp \mdot \vec{u} + 
  \vec{u}^* \mdot \Dfxx (\vec{u},\vec{h}_{01}).
\end{equation}

\subsubsection{Pitchfork bifurcation}
Finally, if $\sigma_0 = 0$ and $g_2 = 0$, but $\sigma_1 \neq
0$ and $g_3 \neq 0$, a pitchfork bifurcation is realized at
$\mu = 0$.  Here the amplitude equation becomes
\begin{equation}
  \dot{z} = \sigma_1 \mu z + g_3 z^3,
  \label{eq:Pitch}
\end{equation}
where $\sigma_1$ is given by \eqn{eq:TransUnC} and 
\begin{equation}
  \label{eq:PitchNonC}
  g_3 = \vec{u}^* \mdot \Dfxx (\vec{u},\vec{h}_{20}) +
  \tfrac{1}{6} \vec{u}^* \mdot \Dfxxx (\vec{u},\vec{u},\vec{u}).
\end{equation}
The results derived for these three bifurcations are
summarized in \tab{tab:SaddleTable}, which also exhibits the
equations for the coefficients of the amplitude
transformation. 

\subsection{Bifurcations at $\lambda=1$ for iterated maps}
The results derived above for flows require little
modification in order to give the coefficients of the
equivalent amplitude equations for saddle-node, transcritical,
and pitchfork bifurcations in the iterated map
\eqn{eq:DiscDiff} (at a fixed point where one eigenvalue of
$\vec{J}$ satisfies $\lambda = 1$ and all other eigenvalues
have $\lambda \neq 1$).  In fact, the only necessary
modification is to
\begin{equation}
  \text{\emph{replace}\, }
  \vec{J}
  \text{\, \emph{by}\, }
  \vec{J}-\vec{I}
  \label{eq:RealTrans}
\end{equation}
everywhere in \eqn{eq:RealBifTwo}, \eqn{eq:RealBifThree},
\eqn{eq:RealUnOne}, and \eqn{eq:RealUnTwo}.  For this reason,
we do not show the details of the calculations for the
iterated map here, but simply give the results for the
unfolded amplitude equations:
\begin{subequations}
  \begin{alignat}{2}
    z & \mapsto \sigma_0\mu + z + g_2 z^2  & 
    \qquad & \text{(\emph{saddle-node})},\\
    z & \mapsto (1+\sigma_1\mu)z + g_2 z^2 & 
    \qquad & \text{(\emph{transcritical})},\\
    z & \mapsto (1+\sigma_1\mu)z + g_3 z^3 & 
    \qquad & \text{(\emph{pitchfork})}.
  \end{alignat}
\end{subequations}
Here the coefficients $\sigma_0$, $\sigma_1$, $g_2$, and $g_3$
are given by \eqn{eq:SaddleUnC}, \eqn{eq:TransUnC},
\eqn{eq:SaddleNonC}, and \eqn{eq:PitchNonC} with the linear
equations \eqn{eq:RealBifTwo}, \eqn{eq:RealBifThree},
\eqn{eq:RealUnOne}, and \eqn{eq:RealUnTwo} modified in
accordance with \eqn{eq:RealTrans}.  The results derived for
the saddle-node, the transcritical, and pitchfork bifurcations
for iterated maps are summarized in \tab{tab:SaddleTable}.

\subsection{Hopf bifurcation}
\label{ssec:SubHopf}
We now consider a stationary point of the flow \eqn{eq:OdePar}
at which the linearization $\vec{J}$ has two complex
conjugate, pure imaginary eigenvalues $\lambda_1 =
\cc{\lambda}_2= \I\omega_0$, and all other eigenvalues have
non-zero real parts.  We denote the right and left
eigenvectors of $\vec{J}$ corresponding to $\lambda_1$ by
$\vec{u}_1$ and $\vec{u}_1^*$, and those of $\lambda_2$ by
$\vec{u}_2$ and $\vec{u}_2^*$, as in the general theory ---
and choose $\vec{u}_2 = \cc{\vec{u}}_1$ normalized according
to \eqn{eq:BiOrtho}.  Consequently, we may simplify the
notation by defining $\vec{u} = \vec{u}_1$ and $\vec{u}^* =
\vec{u}_1^*$.  By this choice, the two-dimensional center
subspace is spanned by any linear combinations of $\re\vec{u}$
and $\im\vec{u}$, whereas the motion in the two-dimensional
center manifold is determined by an amplitude equation which,
as a straightforward application of the general theory, is
formulated in terms of the two complex eigenvalues and complex
conjugate amplitudes $z,\cc{z}$ with $\vec{z} = z\vec{u} +
\cc{z\vec{u}} \in \Ec$.

The ordered set of integers $\vec{p}$ now has two components,
$\vec{p} = (p_1, p_2)$.  The resonance condition for any
$\vec{p}$ is satisfied for either of the two components if
\begin{subequations}
  \begin{align}
    p_1\lambda_1 + p_2\lambda_2 &= (p_1 - p_2)\lambda_1 = \lambda_1\\
    \intertext{or}
    p_1\lambda_1 + p_2\lambda_2 &= (p_2 - p_1)\lambda_2 = \lambda_2.    
  \end{align}
\end{subequations}
Thus, the resonance condition is fulfilled for the first
component if $p_1 = p_2 + 1$, and for the second component if
$p_2 = p_1 + 1$ provided that $p_1 + p_2 \geq 2$.  The
amplitude equation therefore takes the form
\begin{equation}
  \dot{z} = \I\omega_0z + g_{101} \mu z + g_{210} \abs{z}^2\!z + 
  g_{320} \abs{z}^4\!z + \dotsb,
  \label{eq:HopfFormal}
\end{equation}
where $g_{ijk} = \vec{u}^* \mdot \vec{g}_{ijk}$.

To derive expressions for the coefficients of
\eqn{eq:HopfFormal}, we first determine the constant vectors
$\vec{f}_\vec{pq}$ by substitution of
\begin{equation}
  \vec{x} = \vec{u}z + \cc{\vec{u}z} + 
  \vec{h}_{200} z^2 + \vec{h}_{110}\abs{z}^2 + 
  \vec{h}_{020} \cc{z}^2 + \vec{h}_{001} \mu + 
  \vec{h}_{101} \mu z+ \vec{h}_{011} \mu \cc{z} + \dotsb
\end{equation}
in \eqn{eq:NonX}.  Collecting separate orders in $z$,
$\cc{z}$, and $\mu$, we find the following contributions for
terms of zero order in $\mu$
\begin{subequations}
  \begin{align}
    \vec{f}_{200} &= \tfrac{1}{2}\Dfxx (\vec{u},\vec{u}),\\
    \vec{f}_{110} &= \Dfxx (\vec{u},\cc{\vec{u}}),\\
    \vec{f}_{300} &= \Dfxx (\vec{u},\vec{h}_{200}) +
    \tfrac{1}{6}\Dfxxx (\vec{u},\vec{u},\vec{u}),\\
    \vec{f}_{210} &= \Dfxx (\vec{u},\vec{h}_{110}) + 
    \Dfxx (\cc{\vec{u}},\vec{h}_{200}) + 
    \tfrac{1}{2}\Dfxxx (\vec{u},\vec{u},\cc{\vec{u}}).
  \end{align}
  \label{eq:RHSHopf}%
\end{subequations}
whereas the terms linear in $\mu$ are
\begin{subequations}
  \begin{align}
    \vec{f}_{001} &= \Dfp,\\
    \label{eq:HopfRHSOne}%
    \vec{f}_{101} &= \vec{F}_{\vec{x}\mu} \mdot \vec{u} + 
      \Dfxx (\vec{u},\vec{h}_{001}),
  \end{align}
\end{subequations}

Notice also that $\BPhi_{\vec{p}0} = \vec{f}_{\vec{p}0}$ at
second and third orders in $\abs{\vec{p}}$ because the first
non-vanishing coefficients $\vec{g}_{\vec{p}0}$ come at third
order in $\abs{\vec{p}}$.  Although all second order
coefficients $\vec{g}_{\vec{p}0}$ vanish, we must still derive
the coefficients $\vec{h}_{\vec{p}0}$ of the transformation
$\vec{h}$ at second order since these are needed to obtain the
higher order coefficients $\vec{g}_{\vec{p}0}$ as well as for
determining the amplitude transformation.

\vspace{2.5\parsep}
\noindent\textbf{Second order terms in $z$ ($p_1+p_2=2$ and $q=0$)}\\
From \eqn{eq:RHSHopf}, we find 
\begin{subequations}
  \begin{align}
    \BPhi_{200} &= \cc{\BPhi}_{020} = 
    \tfrac{1}{2} \Dfxx (\vec{u}, \vec{u}),\\
    \BPhi_{110} &= \Dfxx (\vec{u}, \cc{\vec{u}}).
  \end{align}
  \label{eq:HopfTwo}%
\end{subequations}

Since there is no resonance at second order, the right-hand
side of \eqn{eq:LinearEqUnA} is simply $-\BPhi_{\vec{p}0}$,
and we get using $\lambda_1 = \cc{\lambda}_2 = \I\omega_0$
\begin{subequations}
  \begin{align}
    (\vec{J} - 2 \I\omega_0 \vec{I}) \mdot \vec{h}_{200} &=
    -\tfrac{1}{2}\Dfxx (\vec{u}, \vec{u}),\\
    \vec{J} \mdot \vec{h}_{110} &= -\Dfxx (\vec{u},\cc{\vec{u}}),
  \end{align}
  \label{eq:HopfLinTwo}%
\end{subequations}
whereas $\vec{h}_{020}$ is determined by the relation
$\vec{h}_{020} = \cc{\vec{h}}_{200}$.  

\vspace{2.5\parsep}
\noindent\textbf{Third order terms in $z$ ($p_1+p_2=3$ and $q=0$)}\\
Once \eqn{eq:HopfLinTwo} has been solved for the second order
coefficient vectors~$\vec{h}_{\vec{p}0}$, we immediately get
the third order terms $\BPhi_{\vec{p}0}$ explicitly from
\eqn{eq:RHSHopf}
\begin{subequations}
  \begin{align}
    \BPhi_{300} &= 
    \Dfxx (\vec{u},\vec{h}_{200}) +
    \tfrac{1}{6}\Dfxxx (\vec{u},\vec{u}, \vec{u}),\\
    \BPhi_{210} &= \Dfxx (\vec{u},\vec{h}_{110}) + 
    \Dfxx (\cc{\vec{u}},\vec{h}_{200}) + 
    \tfrac{1}{2}\Dfxxx (\vec{u},\vec{u},\cc{\vec{u}}),\\
    \BPhi_{120} &= \cc{\BPhi}_{210}, \quad   
    \BPhi_{030} =  \cc{\BPhi}_{300}.
  \end{align}
  \label{eq:HopfPhiTwo}%
\end{subequations}
The third order nonlinear coefficient $g_{210}$ of the
amplitude equation \eqn{eq:HopfFormal} then becomes
\begin{equation}
  g_{210} = \vec{u}^* \mdot \BPhi_{210} =
  \vec{u}^* \mdot \Dfxx (\vec{u},\vec{h}_{110}) + 
  \vec{u}^* \mdot \Dfxx (\cc{\vec{u}},\vec{h}_{200}) + 
  \tfrac{1}{2}\vec{u}^* \mdot \Dfxxx (\vec{u},\vec{u},\cc{\vec{u}}).
\end{equation}
Notice that expressions like \eqn{eq:HopfPhiTwo} for
$\BPhi_{\vec{p}0}$ can be written down immediately in terms of
the coefficients $\vec{h}_{\vec{p}0}$ and
$\vec{g}_{\vec{p}0}$, but to get explicit expressions,
equations like \eqn{eq:HopfLinTwo} must first be solved for
the lower order coefficients.  In other words --- to get
explicit expressions, one must work order by order starting
with the second order.

For the amplitude equation \eqn{eq:HopfFormal} to third order,
we need go no further.  However, to obtain an explicit
expression for the transformation $\vec{h}(\vec{z})$ to third order
we must solve the linear equations
\begin{subequations}
  \begin{align}
    (\vec{J} - 3 \I\omega_0 \vec{I}) \mdot 
    \vec{h}_{300} &= -\BPhi_{300},\\
    (\vec{J} - \I\omega_0 \vec{I}) \mdot \vec{h}_{210} &= 
    -\vec{Q} \mdot \BPhi_{210},\\
    \vec{u}^* \mdot \vec{h}_{210} &= 0,\\
    \vec{h}_{030} &= \cc{\vec{h}}_{300}, \quad
    \vec{h}_{120}  = \cc{\vec{h}}_{210},
  \end{align}
  \label{eq:HopfLinThird}%
\end{subequations}
where the projection $\vec{Q}$ is defined as 
\begin{equation}
  \vec{Q} \mdot \vec{x} = \vec{x} - (\vec{u}^* \mdot \vec{x})\vec{u}.
\end{equation}

\vspace{2.5\parsep}
\noindent\textbf{Unfolding terms ($p_1=p_2=0$ and $q=1$)}\\
First, the coefficient vector $\vec{h}_{001}$, must be found
by solving the linear equation
\begin{equation}
  \label{eq:HopfUnOne}
  \vec{J} \mdot \vec{h}_{001} = -\Dfp,
\end{equation}
which is non-singular since $\lambda_1 = \cc{\lambda}_2 =
\I\omega_0 \neq 0$.  

\vspace{2.5\parsep}
\noindent\textbf{Unfolding terms ($p_1=1$, $p_2=0$ and $q=1$)}\\
We first calculate $\BPhi_{101}$ as 
\begin{align}
  \BPhi_{101} &= \vec{f}_{101} - 
  2 \vec{h}_{200} (\vec{u}^* \mdot \vec{g}_{001}) -
  2 \vec{h}_{110} (\vec{u}^* \mdot \vec{g}_{001}) \\
  &= \vec{f}_{101} = \Dfxp \mdot \vec{u} + \Dfxx (\vec{u},\vec{h}_{001}),
\end{align}
since $\vec{g}_{001} = \vec{0}$.  Having obtained
$\vec{h}_{001}$, the first non-trivial coefficient of the
unfolded amplitude equation becomes
\begin{equation}
  g_{101} = 
  \vec{u}^* \mdot \BPhi_{101} = 
  \vec{u}^* \mdot \Dfxp \mdot \vec{u} + 
  \vec{u}^* \mdot \Dfxx (\vec{u},\vec{h}_{001}),
\end{equation}
whereas the linear equation which determines the corresponding
coefficient vector $\vec{h}_{101}$ in the transformation
becomes
\begin{subequations}
  \begin{align}
    (\vec{J} - \I\omega_0 \vec{I}) \mdot \vec{h}_{101} &= 
    -\vec{Q} \mdot \BPhi_{101},\\
    \vec{u}^* \mdot \vec{h}_{101} &= 0,\\
    \vec{h}_{011} &= \cc{\vec{h}}_{101},
  \end{align}
  \label{eq:HopfLinUn}%
\end{subequations}
We now simplify the notation using the fact that one index, $p
= p_1 + p_2$ say, is sufficient to describe the resonant terms
since $p_1=(p+1)/2$ and $p_2=(p-1)/2$ for the first component.
Setting $\sigma_1 = g_{101}$ and $g_3 = g_{210}$, we finally
conclude that the unfolded amplitude equation becomes
\begin{equation}
  \dot{z} = (\I\omega_0 + \sigma_1 \mu)z + g_3 z\abs{z}^2,
  \label{eq:HopfAmpl}
\end{equation}
with
\begin{subequations}
  \begin{align}
    \label{eq:HopfNonC}
    g_3 &= \vec{u}^* \mdot \Dfxx (\vec{u},\vec{h}_{110}) + 
    \vec{u}^* \mdot \Dfxx (\cc{\vec{u}},\vec{h}_{200}) + 
    \tfrac{1}{2} \vec{u}^* \mdot 
    \Dfxxx (\vec{u},\vec{u},\cc{\vec{u}}),\\
    \sigma_1 &= \vec{u}^* \mdot \Dfxp \mdot \vec{u} + 
    \vec{u}^* \mdot \Dfxx (\vec{u},\vec{h}_{001}).
    \label{eq:HopfUnC}
  \end{align}
\end{subequations}
The results derived for the Hopf bifurcation for flows are
summarized in \tab{tab:HopfTable}.

\subsection{Neimark-Sacker bifurcation}
The equivalent situation where an iterated map
\eqn{eq:DiscDiff} exhibits a Neimark-Sacker bifurcation
corresponds to a fixed point where two complex eigenvalues of
$\vec{J}$ satisfy $\lambda_1 = \cc{\lambda}_2 =
\e^{\I\!\theta_0}$ ($\abs{\lambda_1} = \abs{\lambda_2} = 1$).
For simplicity, we exclude the strong resonance cases where
$\theta_0 = \tfrac{2\pi}{k}$ for $k = 1$, $2$, $3$, or $4$.
The unfolded amplitude equation and the associated
coefficients for this problem is easily obtained by only a few
modifications of the results derived above for the Hopf
bifurcation for flows.  The only required modification is to
\begin{equation}
  \text{\emph{replace}\; }
  \quad
  (\vec{J}-\I k \omega_0\vec{I})
  \quad
  \text{\; \emph{by}\; }
  \quad
  (\vec{J}-\e^{\I\!k\theta_0}\vec{I})
  \quad
  \label{eq:NeimarkTrans}
\end{equation}
everywhere in \eqn{eq:HopfLinTwo}, \eqn{eq:HopfLinThird},
\eqn{eq:HopfUnOne}, and \eqn{eq:HopfLinUn}.  We shall not show
the details of the remaining calculations here but merely
present the main result for the amplitude equation for the
Neimark-Sacker bifurcation, namely
\begin{equation}
  z \mapsto (\e^{\I\! \theta_0} + \sigma_1 \mu)z + 
  g_3 z\abs{z}^2,
  \label{eq:HopfDiscAmpl}
\end{equation}
with the coefficients $g_3$ and $\sigma_1$ given by
\eqn{eq:HopfNonC} and \eqn{eq:HopfUnC} respectively with the
linear equations \eqn{eq:HopfLinTwo}, \eqn{eq:HopfLinThird},
and \eqn{eq:HopfUnOne} modified according to
\eqn{eq:NeimarkTrans}.  The results for the amplitude equation
for the Neimark-Sacker bifurcation are summarized in
\tab{tab:HopfTable}.

\subsection{Period doubling bifurcation}
\label{ssec:PerDouble}
To illustrate the principles of deriving amplitude equations
for iterated maps, we consider a fixed point solution $\vec{x}
= \vec{0}$ of \eqn{eq:DiscDiffUn} where one eigenvalue of
$\vec{J}$ satisfies $\lambda = -1$ at $\mu = 0$ whereas all
other eigenvalues are assumed to have modulus different from
unity.  The right and left eigenvectors are denoted by
$\vec{u}$ and $\vec{u}^*$ respectively corresponding to a
one-dimensional center manifold described by a single real
amplitude $z$ and a simple index $p$.

The resonance condition \eqn{eq:DiscResonance} is simply
$\lambda = \lambda^p$, which is satisfied for all odd $p \geq
3$.  Correspondingly, the amplitude equation
\eqn{eq:DiscAmplEqnFinal} takes the form
\begin{equation}
  z \mapsto (g_{11}\mu - 1)z + g_{30} z^3 + g_{50} z^5 + \dotsb,
  \label{eq:PerAmpl}
\end{equation}
with $g_{pq} = \vec{u}^* \mdot \vec{g}_{pq}$.

To determine relations for the coefficients of
\eqn{eq:PerAmpl}, we first insert
\begin{equation}
  \vec{x} = \vec{u} z + \vec{h}_{20} z^2 + \vec{h}_{30} z^3 +
  \vec{h}_{01}\mu + \vec{h}_{11}\mu z + \dotsb
  \label{eq:PerParamUn}
\end{equation}
in \eqn{eq:NonX} and find the following expressions for
$\vec{f}_{20}$, $\vec{f}_{30}$, $\vec{f}_{01}$, and
$\vec{f}_{11}$
\begin{subequations}
  \begin{align}
    \vec{f}_{20} &= \tfrac{1}{2} \Dfxx (\vec{u},\vec{u}),\\
    \vec{f}_{30} &= \Dfxx (\vec{u},\vec{h}_{20})
    + \tfrac{1}{6} \Dfxxx (\vec{u},\vec{u},\vec{u}),\\
    \vec{f}_{01} &= \Dfp,\\
    \vec{f}_{11} &= \vec{F}_{\vec{x}\mu} \mdot \vec{u} + 
    \Dfxx (\vec{u},\vec{h}_{01}).
  \end{align}
  \label{eq:PerBifThree}%
\end{subequations}

Now, to derive expressions for the constant vectors
$\BPhi_{p0}$ defined implicitly by
\eqnto{eq:DiscInter2}{eq:DiscPhiDefUn}, we note that
$\vec{u}^* \mdot \vec{g}_{20} = 0$, and conclude that
$\BPsi_{p0} = \vec{0}$ for $p \leq 3$ since $\vec{u}^* \mdot
\vec{g}_{p0}$ first enters \eqn{eq:ProdExp} at fourth order.
This therefore implies $\BPhi_{20} = \vec{f}_{20}$ and
$\BPhi_{30} = \vec{f}_{30}$.

\vspace{2.5\parsep}
\noindent\textbf{Second order terms in $z$ ($p=2$ and $q=0$)}\\
To find $\vec{h}_{20}$, we have 
\begin{equation}
  \BPhi_{20} = \tfrac{1}{2} \Dfxx (\vec{u},\vec{u}).
  \label{eq:PerParam}
\end{equation}
There is no resonance at second order, but we must still solve
the regular linear system
\begin{equation}
  (\vec{J} - \vec{I}) \mdot \vec{h}_{20}
  = -\tfrac{1}{2} \Dfxx (\vec{u},\vec{u}),
\end{equation}
for $\vec{h}_{20}$ in order to obtain an explicit expression
for $\BPhi_{30}$.

\vspace{2.5\parsep}
\noindent\textbf{Third order terms in $z$ ($p=3$ and $q=0$)}\\
For $\BPhi_{30}$, we have
\begin{equation}
  \BPhi_{30} = \Dfxx (\vec{u},\vec{h}_{20})
  + \tfrac{1}{6} \Dfxxx (\vec{u},\vec{u},\vec{u}),
\end{equation}
so once the linear equation for $\vec{h}_{20}$ has been
solved, we may determine the resonant third order coefficient
in \eqn{eq:PerAmpl} as
\begin{equation}
  g_{30} = \vec{u}^* \mdot \BPhi_{30} =
  \vec{u}^* \mdot 
  \bigl(
    \Dfxx (\vec{u},\vec{h}_{20}) + 
    \tfrac{1}{6} \Dfxxx (\vec{u},\vec{u},\vec{u})
  \bigr).
\end{equation}
The third order term $\vec{h}_{30}$ of the transformation
$\vec{h}(\vec{z})$ can then be found by solving the equations
\begin{subequations}
  \begin{align}
    (\vec{J} + \vec{I}) \mdot \vec{h}_{30} &=
    -\vec{Q} \mdot \BPhi_{30},\\
    \vec{u}^* \mdot \vec{h}_{30} &= 0,
  \end{align}  
\end{subequations}
where the projection $\vec{Q}$ is defined in \eqn{eq:ProjDef},
\emph{i.e.}\ $\vec{Q} \mdot \vec{x} = \vec{x} - (\vec{u}^*
\mdot \vec{x}) \vec{u}$.

\vspace{2.5\parsep}
\noindent\textbf{Unfolding terms ($p=0,1$ and $q=1$)}\\
To find $\BPhi_{01}$ and $\BPhi_{11}$, we use $\BPhi_{pq} =
\vec{f}_{pq} - \BPsi_{pq}$ and take $\vec{f}_{pq}$ from
\eqn{eq:PerBifThree}.  To determine any contribution from
$\BPsi_{pq}$, we substitute the right-hand side of
\eqn{eq:PerAmpl} for $z$ in the expansion of
$\vec{h}(z,\mu)$.  Now, substract $\vec{h}(-z,\mu)$ and order
the rest according to $(p,q)$.  The coefficient of $z^p \mu^q$
is then $\BPhi_{pq}$.  We may than conclude for the period
doubling bifurcation that $\BPhi_{01} = \BPhi_{11} =
\vec{0}$.  Consequently, 
\begin{subequations}
  \begin{align}
    \BPhi_{01} &= \Dfp,\\
    \BPhi_{11} &= \vec{F}_{\vec{x}\mu} \mdot \vec{u} + 
    \Dfxx (\vec{u},\vec{h}_{01}).
  \end{align}
\end{subequations}
To determine the coefficient vector $\vec{h}_{01}$, we must
solve the linear system
\begin{equation}
  (\vec{J}-\vec{I}) \mdot \vec{h}_{01} = - \Dfp,
  \label{eq:PerDoubleOne}
\end{equation}
which always is regular since $\vec{J}$ by assumption has no
eigenvalue $\lambda = 1$.  Having obtained $\vec{h}_{01}$, we
find the non-trivial contribution to the unfolding of the
amplitude equation as
\begin{equation}
  \vec{u}^* \mdot \vec{g}_{11} = 
  \vec{u}^* \mdot \Dfxp \mdot \vec{u} + 
  \vec{u}^* \mdot \Dfxx (\vec{u},\vec{h}_{01}),
\end{equation}
and the coefficient vector $\vec{h}_{11}$ by solving the
linear system of equations
\begin{subequations}
  \begin{align}
    (\vec{J} + \vec{I}) \mdot \vec{h}_{11} &=
    -\vec{Q} \mdot 
    \bigl(
      \vec{F}_{\vec{x}\mu} \mdot \vec{u} + 
      \Dfxx (\vec{u},\vec{h}_{01})
    \bigl),\\
    \vec{u}^* \mdot \vec{h}_{11} &= 0.
  \end{align}
\end{subequations}

In conclusion, we arrive at the following unfolded amplitude
equation for the period doubling bifurcation
\begin{equation}
  z \mapsto (\sigma_1 \mu - 1)z + g_3 z^3,
  \label{eq:PerAmplUn}
\end{equation}
where
\begin{align}
  \label{eq:PerNonC}
  g_3 &= \vec{u}^* \mdot \Dfxx (\vec{u},\vec{h}_{20}) + 
  \tfrac{1}{6} \vec{u}^* \mdot \Dfxxx (\vec{u},\vec{u},\vec{u}),\\
  \label{eq:PerUnC}
  \sigma_1 &= \vec{u}^* \mdot \Dfxp \mdot \vec{u} + 
  \vec{u}^* \mdot \Dfxx (\vec{u},\vec{h}_{01}).
\end{align}
The results for the period doubling bifurcation are summarized
in \tab{tab:DoubleTable}.

\section{Scalings and proper amplitude equations}
\label{sec:Scalings}
The amplitude of a harmonic oscillation is half the difference
between the maximum and minimum values of the oscillating
function.  This basic concept of an amplitude can be
generalized to a set of (possibly complex) Fourier amplitudes
for anharmonic periodic oscillations.  For the aperiodic
(transient) oscillations that can occur near a Hopf
bifurcation, we can introduce a slowly varying, complex
amplitude $w$ related to $z$ of \eqn{eq:HopfAmpl} by
extracting the time dependence associated with harmonic
oscillations with (vanishingly) small amplitude at the
bifurcation point.  Specifically, we can extract a phase
factor $\e^{\I\! \omega_0 t}$ from $z(t)$ to get a time
dependent amplitude modulating the ``basic'' oscillations
described by the phase $\e^{\I\!\omega_0 t}$.

We shall introduce the notion of a ``proper'' amplitude to
account for this and similar slow modulations.  At the same
time, it is convenient to account also for the important
scalings characteristic of a Hopf bifurcation.  Thus, the
amplitude of the limit cycle oscillations grows as
$\sqrt{\mu}$ for small $\mu$, and the characteristic times in
a vicinity of the stationary point as well as near the limit
cycle are both proportional to $\mu$ close enough to the
bifurcation point.

We therefore introduce a proper (scaled) amplitude $w(\tau)$
as a function of a scaled time $\tau$ by the definition
\begin{align}
  w(\tau) &= \frac{1}{\sqrt{\mu}} \exp (-\I\omega_0 t) z(t),\\
  \tau &= \mu t.
\end{align}
Substitution of these two expressions in \eqn{eq:HopfAmpl}
then leads to the proper amplitude equation for $w(\tau)$
\begin{equation}
  \frac{dw}{d\tau} = \sigma_1 w + g_3 w\abs{w}^2,
  \label{eq:HopfAmplScaled}
\end{equation}
which traditionally is called the Stuart-Landau
equation~\cite{LandauTurb,Stuart}.  It has the important
property of being independent of the distance from the
bifurcation point (which of course only applies at the lowest
non-trivial orders in $z$ and $\mu$ of the original amplitude
equation).  So the Stuart-Landau equation is universal for any
flow exhibiting a Hopf bifurcation --- characterized by just
two parameters $\sigma_1$ and $g_3$.  Note that the scalings
leading to \eqn{eq:HopfAmplScaled} are dictated by the form of
\eqn{eq:HopfAmpl}.  The Stuart-Landau equation
\eqn{eq:HopfAmplScaled} can be derived from \eqn{eq:HopfAmpl}
without any knowledge of the properties of the Hopf
bifurcation other than those contained in \eqn{eq:HopfAmpl}.

Our remarks leading to \eqn{eq:HopfAmplScaled} are meant
solely as motivation of the idea of scaling and are not needed
for the derivation itself.  On the contrary, we can infer the
scaling properties for the limit cycle and the critical
slowing down from the amplitude equation.  This feature
distinguishes the present derivation of (proper) amplitude
equations from those based on multiple time scales; in the
latter case the scalings are basic assumptions of the actual
derivation of the amplitude equation whereas these follow
naturally from the derivations presented here.

For non-oscillatory modes, the amplitude equations at lowest
nonlinear order can also be scaled to yield a proper amplitude
equation which also will be universal for any flow exhibiting
the bifurcation in question.  In fact, similar proper
amplitude equations are easily derived for the saddle-node,
the transcritical, and the pitchfork bifurcation.  For the
equivalent amplitude equations of an iterated map, one can
show that the scaling $t = \mu\tau$ is replaced by an explicit
factor $\mu$ in the scaled amplitude equation.

\section{Interpretation of the unfoldings}
\label{sec:UnInt}
The significance of the description of the motion in the
center manifold in terms of an amplitude equation is obvious
from the geometrical interpretation illustrated
in~\fig{fig:AmpMap}.  But the meaning of the unfolding is
perhaps not so clear.

To understand terms like the coefficient vector
$\vec{h}_{\vec{pq}}$ in \eqn{eq:LinearEqUn} for $\abs{\vec{q}}
\neq 0$, we look first at the case $\abs{\vec{p}} = 0$,
$\abs{\vec{q}} = 1$ corresponding to
\eqnto{eq:UnfoldZero}{eq:UnfoldZeroG}.  These equations are
associated with the change of the stationary point
$\stab{x}(\bmu)$ with the parameter $\bmu$ as described by the
equation
\begin{equation}
  \vec{F}
  \bigl(
    \stab{x}(\bmu),\bmu
  \bigr) = \vec{0}.
\end{equation}
The change from $(\stab{x},\bmu) = (\vec{0},\vec{0})$ is to
first order in $\bmu$ determined by the equation
\begin{equation}
  \difd{\vec{F}}{\bmu} \mdot \bmu  =
  \bigl(
    \vec{J} \mdot \difp{\stab{x}}{\bmu} + \Dfp
  \bigr) \mdot \bmu = \vec{0}
  \label{eq:StatVar}
\end{equation}
in which all derivatives are evaluated at $(\vec{x},\bmu) =
(\vec{0},\vec{0})$.  For finite $\bmu$, the expression inside
the parentheses in \eqn{eq:StatVar} must vanish, and we find
\begin{equation}
    \vec{J} \mdot \difp{\stab{x}}{\bmu} = -\Dfp,
\end{equation}
which corresponds to \eqn{eq:UnfoldZero} for the $k$'th
component of $\bmu$.

Thus the component of $\vec{h}_{\vec{0}\bdl_{\!k}}$ in the
range of $\vec{J}$ determines the change in $\R^n$ of the
stationary point $\stab{x}(\bmu)$ as
\begin{equation}
  \stab{x}(\bmu) = \sum_{k=1}^s \vec{h}_{\vec{0}\bdl_{\!k}} \mu_k.
  \label{eq:StatZero}
\end{equation}
A change with $\bmu$ of the motion in the null\-space of
$\vec{J}$ is determined by the contribution to the amplitude
equation \eqn{eq:AmplUnfold} provided the derivative $\Dfp$
has a non-vanishing component in the null\-space of $\vec{J}$.
In contrast, the transformation $\vec{h}$ will have no
component in the nullspace.

We see the effects of the $\abs{\vec{p}} = 0$, $\abs{\vec{q}}
= 1$ contributions to the unfoldings for the single zero
eigenvalue in Section~\ref{ssec:RealBif} and for the Hopf
bifurcation in Section~\ref{ssec:SubHopf}.  Thus, the change
of the stationary point is given to first order in $\mu$ by
\eqn{eq:RealUnOne} and \eqn{eq:HopfUnOne}.  In this case,
there is only a contribution to the amplitude equation for the
saddle-node bifurcation where $\Dfp$ has a non-vanishing
component in the nullspace of $\vec{J}$.  Here the
non-hyperbolic stationary point splits in two which cannot be
accounted for as a smooth displacement of the stationary point
under the unfolding --- a term must therefore appear in the
amplitude equation.  For the transcritical, pitchfork, and
Hopf bifurcations, on the other hand, the change of the
stationary point with $\mu$ can simply be accounted for by the
smooth transformation $\vec{h}(\vec{z},\bmu)$.

At the order $\abs{\vec{p}} = 1$, $\abs{\vec{q}} = 1$ the
unfolding can be understood in terms of perturbation theory
for the eigenvalue problem.  Consider the change of the
Jacobian matrix at the stationary point $\stab{x}$ to first
order in $\bmu$
\begin{equation}
  \vec{J}(\bmu) = \vec{J} + 
  \vec{F}_{\vec{x}\bmu}(\cdot,\bmu) + \Dfxx(\cdot,\stab{x}),
  \label{eq:PertEig}
\end{equation}
in which $\stab{x}$ (to first order in $\bmu$) is given by
\eqn{eq:StatZero}.  For short we write \eqn{eq:PertEig} as
\begin{equation}
  \vec{J}(\bmu) = \vec{J} + \delta\vec{J}
\end{equation}
and consider the eigenvalue problem 
\begin{equation}
  (\vec{J}+\delta\vec{J}) \mdot \; (\vec{u}_j + \delta\vec{u}_j) =
  (\lambda_j + \delta\lambda_j) 
  (\vec{u}_j + \delta\vec{u}_j).
\end{equation}
To lowest order (linear in $\bmu$), we find from perturbation
theory \cite{Kato} that the change $\delta\lambda_j$ of the
eigenvalue $\lambda_j$ is
\begin{equation}
  \delta\lambda_j = 
  \vec{u}^*_j \mdot \delta\vec{J} \mdot \vec{u}_j,
\end{equation}
whereas the change $\delta\vec{u}_j$ of the eigenvector
$\vec{u}_j$ is determined by the linear equation
\begin{equation}
  \begin{split}
    (\vec{J} - \lambda_j \vec{I}) \mdot \delta\vec{u}_j &=
    -(\vec{I} - \vec{u}_j\vec{u}_j^*) \mdot \delta\vec{J}
    \mdot \vec{u}_j,\\ \vec{u}_j^* \mdot \delta\vec{u}_j &= 0.
  \end{split}
  \label{eq:PertOne}
\end{equation}
From \eqn{eq:PertEig}, we find for $\delta\vec{J} \mdot
\vec{u}_j$
\begin{equation}
  \delta\vec{J} \mdot \vec{u}_j = 
  \Dfxp(\vec{u}_j,\bmu) + \Dfxx(\vec{u}_j,\stab{x}).
  \label{eq:PertTwo}
\end{equation}

By comparison of \eqn{eq:PertOne} and \eqn{eq:PertTwo} with
\eqn{eq:UnfoldOne}, we identify $\vec{h}_{\bdl_{\!j} \!
\bdl_{\!k}}$ as the contribution to the change of the $j$'th
eigenvector per unit change of $\mu_k$.  Similarly, we
identify the coefficient $\vec{u}_j^* \mdot
\vec{g}_{\bdl_{\!j} \! \bdl_{\!k}}$ in the unfolded amplitude
equation as the derivative of the $j$'th eigenvalue with
respect to $\mu_k$.

The effects of the $\abs{\vec{p}} = 1$, $\abs{\vec{q}} = 1$
contributions can be inferred directly from the specific
bifurcations studied in \sects{ssec:RealBif}{ssec:SubHopf}.
For a scalar parameter $\mu$, we have $\BPhi_{\bdl_{\!j} \!
\bdl_{\!k}} = \BPhi_{\bdl_{\!j}1}$ implying that
\eqn{eq:PertTwo} reduces to
\begin{equation}
  \delta\vec{J} \mdot \vec{u}_j = 
  \Dfxp \mdot \vec{u}_j + \Dfxx(\vec{u}_j,\vec{h}_{\vec{0}1}).
\end{equation}
The right-hand side of this expression has the same form as
\eqn{eq:SaddleRHSOne} and \eqn{eq:HopfRHSOne} for a single
zero eigenvalue and for a Hopf bifurcation respectively.  For
a scalar parameter, we also note that the change of the
stationary point is simply $\vec{h}_{\vec{0}1}\mu$
corresponding to first order terms $\vec{h}_{01}\mu$ and
$\vec{h}_{001}\mu$ in \sects{ssec:RealBif}{ssec:SubHopf}
respectively.

We can summarize these results for one critical component with
eigenvalue $\lambda$, right eigenvector $\vec{u}$ and
amplitude $z$ in simplified notation (disregarding other
critical modes) by grouping terms as follows
\begin{subequations}
  \begin{align}
    \dot{z} &=       g_{01}\mu + (\lambda + g_{11}\mu)z + \dotsb,\\
    \vec{x} &= \vec{h}_{01}\mu + (\vec{u} + \vec{h}_{11}\mu)z + \dotsb,
  \end{align}
  \label{eq:Int}%
\end{subequations}
where '$\dotsb$' indicates linear contributions from other
critical components plus nonlinear terms.

Thus, the result \eqn{eq:Int} partly accounts for the change
of the stationary point to $\vec{h}_{01}\mu$ and changed
eigenvalue $\lambda + g_{11}\mu$ and eigenvector $\vec{u} +
\vec{h}_{11}\mu$ evaluated at the new stationary point.  It
seems reasonably to assume that other terms can be interpreted
as corrections changing the amplitude, but we shall not pursue
the interpretation any further.

The interpretation indicated above suggest that for finite
$\bmu$ one may obtain an efficient description of the dynamics
by using the formulas obtained for the center manifold using
the (nearly critical) eigenvalues and eigenvectors as well as
all coefficients calculated from the vector field at
$\stab{x}(\bmu)$ for the finite value of $\bmu$.

In the important special case where no eigenvalues have
positive real parts at the bifurcation and all the critical
eigenvalues and no others acquire positive real parts at the
finite $\bmu$, we can appeal to the stable manifold theorem
and use essentially the same line of reasoning as in
\sect{sec:Flows} for the description of the motion in the
unstable manifold.  We may still classify the terms to include
in the amplitude equation and those to include in the
associated transformation according to whether or not the
resonance condition is satisfied \emph{at the bifurcation
point}.  Although it is in general possible to eliminate all
nonlinear terms from the amplitude equation, one should not do
so, but instead include precisely those that appear at the
bifurcation.  Then the amplitude equation and transformation
change smoothly with $\bmu$, and the formulas derived here for
the center manifold can be applied to the unstable manifold
with all quantities calculated at $\stab{x}(\bmu)$ for finite
$\bmu$.

\section{Physical relevance of the amplitude equation} 
\label{sec:PhysRela}
An amplitude equation describes the kind of dynamical behavior
(\emph{e.g.}\ chaos) that can be observed in any system to
which it applies.  It can therefore be studied in its own
right as a representative of a large class of dynamical
systems.

For a specific system, the amplitude equation has its
coefficients fixed by the physical system.  Some or all of
these coefficients can often be determined from a summary
knowledge of the system, such as characteristic (invariant)
exponents.  Furthermore, since the actual (physical) system is
topologically equivalent to that determined by the amplitude
equation, the number of solutions that branch off at the
bifurcation point and the stability of the bifurcating
solutions (the super- and subcritical nature of the
bifurcation) are additional examples of characteristics that
are determined directly by the properties of the amplitude
equation.

However, to describe a real physical system in more detail in
terms of amplitudes, the amplitude equation is not sufficient.
The transformation \eqn{eq:PhysRela} between the description
in terms of amplitudes and the actual physical variables is
essential.  Fortunately, that transformation is obtained
together with the amplitude equation with little extra work,
since the coefficients $\vec{h}_\vec{pq}$ of the
transformation must be calculated in order to determine the
amplitude equation anyway.

For a given amplitude equation, we may therefore study the
dynamics by either analytic or numerical methods and then use
\eqn{eq:PhysRela} to transform the dynamics back to the
physical space of the original problem.  For all the amplitude
equations derived in \sect{sec:GenExamp}, an analytic approach
is possible.  The calculation of the limit sets of these
amplitude equations is straightforward.  The results and their
relation to the physical space is summarized in
\tabto{tab:SaddleTable}{tab:DoubleTable} and \tab{tab:Phys}
respectively.

The advantage of using an amplitude approach to a physical
problem is that the essential dynamics usually takes place in
a subspace (the center subspace of amplitudes) of much lower
dimension than the full dimension of the dynamical system.  So
whether one has an experiment or a model, it is often
advantageous to solve or analyze the problem in terms of
amplitudes.  For such work, a quantitative approach like that
described in the present paper is necessary.  As an example we
mention a quantitative experimental analysis of oscillatory
chemical reactions near a supercritical Hopf bifurcation
relying on the Stuart-Landau
equation~(\ref{eq:HopfAmplScaled}) and the basic geometry of
the amplitude transformation~\cite{GenQuench,PgsCgle} and a
study of chemical waves in a complex chemical
reaction-diffusion system~\cite{Ipsen96}.

\section{Summary and discussion}
\label{sec:Summary}
The main result of this paper is a general and systematic
method of deriving amplitude equations (or normal forms) for
flows and iterated maps near a local bifurcation.  In
addition, the method also provides the transformations that
connect the amplitude description with the ``real world''
described in terms of appropriate physical variables, the
amplitude transformation.  In particular, the method provides
an explicit recurrence relation \eqn{eq:GSolveUnfold} with
\eqn{eq:PhiDefUnfold} for flows, and \eqn{eq:DiscPhiDefUn} for
iterated maps in the case of semisimple critical eigenvalues. 

The method is completely transparent: the amplitudes
representing the instantaneous state of a system and the
amplitude transformation to a physical description of the
state have simple geometric interpretations.  The use of the
method for a given type of bifurcation is straightforward and
unambiguous, and can in principle be carried out to any
desired order of nonlinearity.

The method is efficient in terms of necessary computational
efforts as will become clear from a summary of the steps of
the method which we now present: We first have to find the
eigenvalues $\lambda_i$ and the associated eigenvectors
$\vec{u}_i$ and $\vec{u}_i^*$ satisfying the condition
\mbox{$\re\lambda_i = 0$} (for flows) or
\mbox{$\abs{\lambda_i} = 1$} (for iterated maps).  From this,
we immediately get the structure of the amplitude equation and
the amplitude transformation $\vec{h}(\vec{z},\bmu)$ through a
classification of terms $\vec{u}_i^* \mdot \vec{g}_\vec{pq}$
and $\vec{u}_i^* \mdot \vec{h}_\vec{pq}$ depending on whether
or not the resonance condition
\begin{subequations}
  \begin{alignat}{2}
     \sum_{j=1}^r p_j \lambda_j &= \lambda_i & 
     \qquad & \text{(for flows)},\\
     \prod_{j=1}^r \lambda_j^{p_j} &= \lambda_i & 
     \qquad & \text{(for maps)}
  \end{alignat}
\end{subequations}
is satisfied for the $i$'th component (amplitude $z_i$) of the
term indicated by the order $(\vec{p},\vec{q})$.  The
amplitude equation includes all resonant terms and no others,
whereas the amplitude transformation contains non-resonant
terms only.

The next step is to calculate the constant vectors
$\BPhi_\vec{pq}$ in terms of the Taylor coefficients
$\vec{f}_\vec{pq}$, $\vec{g}_\vec{pq}$, and
$\vec{h}_\vec{pq}$.  This step is straightforward.  For simple
bifurcations (as considered in this paper), the necessary
calculations are easily done by hand, whereas more complex
problems of higher codimension is a simple task for software
tools such as Mathematica and Maple~\cite{SW96,Maple}.

The coefficients $\vec{g}_\vec{pq}$ (for the amplitude
equation) and $\vec{h}_\vec{pq}$ (for the amplitude
transformation) are now obtained order by order in an
iterative procedure.  At any order $(\vec{p},\vec{q})$, any
non-vanishing component $\vec{u}_i^* \mdot \vec{g}_\vec{pq}$ in
the amplitude equation is given explicitly by
\begin{equation}
  \vec{u}_i^* \mdot \vec{g}_\vec{pq} = 
  \vec{u}_i^* \mdot \BPhi_\vec{pq}.
\end{equation}
This term only appears if the resonance condition is
fulfilled.  To obtain the coefficient $\vec{h}_\vec{pq}$, one
has to solve the system of linear equations
\begin{subequations}
  \label{eq:Sum}
  \begin{alignat}{2}
    \label{eq:SumA}
    \Bigl(
      \vec{J}  - \sum_{j=1}^r p_j \lambda_j \vec{I}
    \Bigr) \mdot  
    \vec{h}_\vec{pq} &= - \vec{Q}_\vec{p} \mdot \BPhi_\vec{pq} &
    \qquad & \text{(for flows)},\\
    \label{eq:SumB}
    \Bigl(
      \vec{J}  - \prod_{j=1}^r \lambda_j^{p_j} \vec{I}
    \Bigr) \mdot  
    \vec{h}_\vec{pq} &= - \vec{Q}_\vec{p} \mdot \BPhi_\vec{pq} &
    \qquad & \text{(for maps)}
  \end{alignat}
\end{subequations}
determining the non-resonant components of $\vec{h}_\vec{pq}$,
with the auxiliary condition
\begin{equation}
  \label{eq:SumC}
  \vec{R}_\vec{p} \mdot 
  \vec{h}_\vec{pq} = \vec{0},
\end{equation}
which ensures that the resonant components of
$\vec{h}_\vec{pq}$ vanish.  \Eqn{eq:SumA} or (\ref{eq:SumB})
together with \eqn{eq:SumC} determine $\vec{h}_\vec{pq}$
completely.

This is all that is needed to formulate a given physical
problem in terms of an amplitude equation.  Notice carefully
that the above procedure does \emph{not} require that the
Jacobian matrix has been transformed to block form --- a
feature that greatly simplifies the computations.  The only
non-trivial parts of the computations are the determination of
the right and left eigenvectors associated with the
bifurcating eigenvalues and the solution of the systems of
linear equations \eqn{eq:Sum}.

Another significant result of the present paper is the
compilation of general results for several types of
bifurcations as summarized in
\tabto{tab:SaddleTable}{tab:Phys}.  This part of the work
serves two important purposes.  First we use it to demonstrate
how the general procedure (sketched above) works in practice
for flows and iterated maps.  Secondly, it provides explicit
expressions that can be used immediately for any specific
problem.  Such actual use of the tables is illustrated in
Appendix~\ref{sec:Examples} where the calculation of amplitude
equations and associated transformations is demonstrated for a
few specific problems, using the formulas of the tables.

As can be seen from the specific bifurcations discussed in
\sect{sec:GenExamp} and the examples of the Appendix, the
machinery is quite easy to use despite the array of expansions
involved.  In practice, these expansions are finite or can be
truncated so that only few terms need be considered.  Usually,
only a few critical modes are involved.  Thus the amplitude
equation has few components, and in general one needs only
include its first few terms to get a satisfactory description
of the dynamics.  The associated amplitude transformation can
usually be truncated after the highest order included in the
amplitude equation.  We also emphasize that considerable
simplification occurs in many applications as a result of the
form of the vector field or map.  For example, this is the
case in chemical kinetics based on elementary reactions, which
are at most of the second order.  Consequently, the Taylor
expansion \eqn{eq:NonlinExp} terminates at low order because
higher order derivatives vanish.  As mentioned previously, if
the derivations by hand become too laborious, one can always
carry it through by resorting to software package like
Mathematica or Maple which are capable of performing symbolic
computations

The main application of the results of the present paper is
for determining the time evolution of dynamical systems near a
local bifurcation.  However, the derivation of an amplitude
equation also provides an explicit expression for the center
manifold in the simplest possible way.

In addition, the formulas of \tabto{tab:SaddleTable}{tab:Phys}
provide a general tool for determining numerical estimates of
branchings of stationary, fixed points, and periodic solutions
of flows and maps when a bifurcation is traversed.  Such
estimates are typically needed in continuation problems as
implemented in the numerical software packages CONT \cite[pp.\
302--361]{Cont,Marek91} and AUTO \cite{Auto97}.  Estimates
based on the solutions of amplitude equations are more
accurate than those traditionally used in terms of tangents of
the intersecting branches at the bifurcation point
\cite{Marek83}.  These do not include the curvature of the
branches at the bifurcation and equally important --- they do
not determine the stability character (as super- or
subcritical) of the bifurcation point.  All of these issues
are generally important when calculating delicate solution and
bifurcation diagrams.  For details regarding the numerical
location of bifurcation points of the type discussed in this
paper, we refer to \cite{Marek83,HK87,WJB91} and references
therein.

Notice also that the method presented here is generally valid
also for stable and unstable manifolds.  Except for accidental
resonance, the differential equation describing the dynamics
on these manifolds can in general be linearized to any order
implying that all terms are non-resonant and therefore
contained in the associated transformation.

As already mentioned in \sects{sec:Flows}{sec:Disc}, we
emphasize that the theory for amplitude equations can be used
directly to perform reductions of flows and iterated maps to
normal form.  Traditional approaches, where the normal form
theorem is applied directly to a given problem, involves the
determination of Lie brackets and solutions of the associated
homological equation.  As noted by Arrowsmith \&
Place~\cite{ArrPla90}, such calculations are often complicated
and rather lengthy.  The approach considered here, however,
determines the normal form and the associated (normal form)
transformation simply by solving systems of linear equations.
From practical experience, we find the approach considered
here much more convenient compared to traditional normal form
reductions.

In this paper, we have worked out explicit formulas for a
number of simple bifurcations.  The derivation of these serve
as an illustration of the general method presented but also
provides a collection of results than can be useful for
applications to specific systems: it may indeed save much
time.

However, the method of deriving amplitude equations and
associated transformations is absolutely straightforward also
for more complicated local bifurcations (of codimension-two
and higher).  We have derived explicit amplitude equations and
transformations for some of these higher bifurcations for
flows and maps with the method described here.  In this
connection, it is natural to treat also bifurcations where the
geometric multiplicity of the critical eigenvalues is smaller
than the algebraic multiplicity.  Such bifurcations can be
treated in terms of generalized eigenvectors by
straightforward modification of the present method.  However,
the general case requires a complicated notation, so we have
not made the extension to generalized center eigenspace in the
present paper.  At the time of writing, we have worked out
explicit expressions for coefficients of amplitude equations
and transformations for a Bautin bifurcation (where a Hopf
bifurcation changes from super- to subcritical and a quintic
term is needed), for a Bogdanov-Takens bifurcation, for a
fold-Hopf bifurcation, and for a ``Hopf-Hopf'' bifurcation
(where two pairs of complex conjugate eigenvalues pass the
imaginary axis) with and without strong resonances.  For
iterated maps, we have worked out the four cases of strong
resonance for the Neimark-Sacker bifurcation.  

In a future paper, we shall show how an amplitude equation for
a fold-Hopf bifurcation can be applied on a quantitative level
to describe a realistic reaction-diffusion model system for
the Belousov-Zhabotinsky reaction under circumstances where
the complex Ginz\-burg-Lan\-dau equation fails entirely.  In
fact, this observation suggests that such codimension-two
descriptions may be much more useful than the codimension
initially suggests.  In particular, that paper will illustrate
how results for amplitude equations and their associated
transformations can be used to obtain a description in
physical terms that may be compared quantitatively with the
full model system and experimental data.

\appendix

\section{Examples}
\label{sec:Examples}
In order to illustrate the theory discussed in the previous
sections and demonstrate the use of the formulas of
\tabto{tab:SaddleTable}{tab:Phys}, we discuss three simple
examples.  They will show the fundamental steps involved in
calculating the amplitude equation for various bifurcations in
actual systems.  We emphasize that the use of amplitude
equations for the particular examples discussed here has been
chosen to illustrate the principles of the method for some
simple bifurcations --- not because it necessarily provides
the computationally shortest path to the solution.

\subsection{The Lorenz equations}
To illustrate the derivation of amplitude equations for
bifurcations involving a single zero eigenvalue, we consider
the Lorenz equations~\cite{Lorenz63}
\begin{equation}
  \label{eq:Lorenz}
  \begin{pmatrix}
    \dot{x}_1 \\
    \dot{x}_2 \\
    \dot{x}_3 
  \end{pmatrix} =
  \begin{pmatrix}
    s (x_2 - x_1)\\
    r x_1 - x_2 - x_1x_3\\
    -b x_3 + x_1x_2
  \end{pmatrix},
\end{equation}
and choose the term $r$ as the bifurcation parameter.  One
easily sees that \eqn{eq:Lorenz} admits the trivial solution
$(0,0,0)^\trp$ with the corresponding Jacobian
\begin{equation}
  \vec{J}(r) = 
  \begin{pmatrix}
    -s &  s &  0 \\
     r & -1 &  0 \\
     0 &  0 & -b 
  \end{pmatrix},
\end{equation}
which has a single real eigenvalue $\lambda = 0$ for $r = 1$.
This implies that the Lorenz system has a one-dimensional
center manifold at the stationary point $(0,0,0)^\trp$ for
$r=1$.  The right and left eigenvectors $\vec{u}$ and
$\vec{u}^*$ of $\vec{J}=\vec{J}(1)$ are
\begin{align}
  \vec{u}   &= (1,1,0)^\trp, \\
  \vec{u}^* &= \frac{1}{1+s}(1,s,0).
\end{align}

\noindent\textbf{Second and third order terms}\\
Since the coefficient $\sigma_0$ in \tab{tab:SaddleTable}
satisfies $\sigma_0 = \vec{u}^* \mdot \vec{F}_r = \vec{u}^*
\mdot \difp{\vec{F}}{r}\bigl|_{r=1} = 0$, we may exclude the
possibility of a saddle-node bifurcation.  Furthermore, since
\begin{equation}
  \Dfxx(\bxi,\bet) = 
  (0, -\xi_1 \eta_3 - \eta_1 \xi_3, \xi_1 \eta_2 + \eta_1 \xi_1)^\trp,
\end{equation}
we may calculate the quadratic coefficient $g_2$ from
\tab{tab:SaddleTable} as
\begin{equation}
  g_2 = \tfrac{1}{2} \vec{u}^* \mdot \Dfxx (\vec{u},\vec{u})
      = \vec{u}^* \mdot \,(0,0,1)^\trp
      = 0,
\end{equation}
which excludes the possibility of a transcritical bifurcation.
In order to determine the third order coefficient $g_3$, we
must first find the coefficient vector $\vec{h}_{20}$.  From
\tab{tab:SaddleTable} we get
\begin{align}
  \vec{J} \mdot \vec{h}_{20} &= -(0,0,1)^\trp,\\
  \vec{u}^* \mdot \vec{h}_{20} &= 0,
\end{align}
implying that $\vec{h}_{20} = (0,0,\frac{1}{b})^\trp$.  Observing
that the Taylor expansion of \eqn{eq:Lorenz} terminates at
second order ($\Dfxxx = \vec{0}$), we may determine the cubic
coefficient $g_3$ from \tab{tab:SaddleTable} as
\begin{equation}
  g_3 = \vec{u}^* \mdot \Dfxx (\vec{u},\vec{h}_{20}) = 
  \vec{u}^* \mdot\, (0,-\frac{1}{b},0)^\trp = -\frac{s}{b(1+s)}.
\end{equation}

\noindent\textbf{Unfolding terms}\\
To find the coefficient $\sigma_1$ in the unfolding of the
amplitude equation associated with pitchfork bifurcation (see
\tab{tab:SaddleTable}), we first observe that $\vec{F}_r =
\vec{0}$ implying that the coefficient vector $\vec{h}_{01}$
vanishes implying that the term $\vec{u}^* \mdot \Dfxx
(\vec{u},\vec{h}_{01})$ in $\sigma_1$ disappears.
Consequently, we get from \tab{tab:SaddleTable}
\begin{equation}
  \sigma_1 = \vec{u}^* \mdot \vec{F}_{\vec{x}r} \mdot \vec{u}
  = \vec{u}^* \mdot 
  \begin{pmatrix}
    0 & 0 & 0\\
    1 & 0 & 0\\
    0 & 0 & 0\\
  \end{pmatrix} \mdot 
  \vec{u}
  = \frac{s}{s + 1}.
\end{equation}
For $s \neq 0$, we have $g_3 \neq 0$, and $\sigma_1 \neq 0$,
and therefore conclude that a pitchfork bifurcation occurs in
the Lorenz system \eqn{eq:Lorenz} at $r = 1$.  If $s > 0$ and
$b > 0$, the bifurcation will be supercritical.

From \tab{tab:SaddleTable}, we now obtain the following
expression for the amplitude equation for the Lorenz system
\eqn{eq:Lorenz}
\begin{equation}
  \dot{z} = 
  \frac{s}{s + 1}
  \Bigl[
    (r - 1) z - \frac{1}{b} z^3
  \Bigr],
  \label{eq:AmplLorenz}
\end{equation}
which admits the non-trivial stationary solutions
\begin{equation}
  z_{\pm} = \pm\sqrt{b(r - 1)}.
\end{equation}

To determine the branching behavior of \eqn{eq:Lorenz} in the
physical $(x_1,x_2,x_3)$-space, we use the amplitude
transformation associated with the pitchfork bifurcation from
\tab{tab:Phys}.  We find
\begin{align}
  \begin{pmatrix}
    x_1 \\
    x_2 \\
    x_3 
  \end{pmatrix}_{\!\!\pm} =
  \pm
  \begin{pmatrix}
    1 \\
    1 \\
    0 
  \end{pmatrix}\sqrt{b(r - 1)} +
  \begin{pmatrix}
    0 \\
    0 \\
    1 
  \end{pmatrix}(r - 1).
\end{align}
We have not shown the calculation of the coefficient vectors
$\vec{h}_{30}$ and $\vec{h}_{11}$, since these accidentally
cancel when inserted into the amplitude transformation.

\subsection{The Gray-Scott model}
As an example of a two-dimensional chemical system that
exhibits a Hopf bifurcation, we consider the Gray-Scott
model~\cite{GS85} for the transformation of a reactant A to a
product B with two intermediate species X$_1$ and X$_2$:
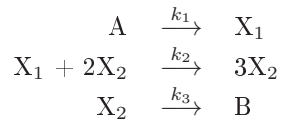
\begin{center}
  \begin{tabular}{rll}
    A               & $\stackrel{k_1}\longrightarrow$ & X$_1$\\
    X$_1$ + 2X$_2$  & $\stackrel{k_2}\longrightarrow$ & 3X$_2$\\
    X$_2$           & $\stackrel{k_3}\longrightarrow$ & B
  \end{tabular}
\end{center}

Introducing dimensionless concentrations by $x_1 =
\sqrt{k_2/k_1}[{\rm X}_1]$ and $x_2 = \sqrt{k_2/k_1}[{\rm
X}_2]$, a dimensionless time by $\tau = k_1t$, and choosing
$k_3 = k_1$ the scaled kinetic equations become

\begin{equation}
  \label{eq:Roessler}
  \begin{pmatrix}
    \dot{x}_1 \\
    \dot{x}_2
  \end{pmatrix} =
  \begin{pmatrix}
    \nu -x_1 x_2^2\\
    x_1 x_2^2 - x_2
  \end{pmatrix},
\end{equation}
where $\nu = \sqrt{k_2/k_1}[\mathrm{A}]$ is used as
bifurcation parameter.  Here we consider the stationary
solution ${\bf x_s} = (\frac{1}{\nu},\nu)$.  The Jacobian of
\eqn{eq:Roessler} at ${\bf x}_s$ is
\begin{equation}
  \vec{J}(\nu) =
  \begin{pmatrix}
    -\nu^2 & -2 \\
    \nu^2 &  1
  \end{pmatrix}
\end{equation}

For $\nu = 1$ the Jacobian $\vec{J}$ has eigenvalues
$\lambda_1 = \overline{\lambda}_2 = \I$ implying that the
system exhibits a Hopf bifurcation.  The first order terms are
given by the right and left complex eigenvectors $\vec{u}$ and
$\vec{u}^*$ associated with $\lambda_1$.  We get
\begin{equation}
  \vec{u} =  (-2, 1 + \I)^\trp 
  \qquad  \text{and} \qquad
  \vec{u}^* = -\frac{1}{4}(1 + \I, 2\I)^\trp.
\end{equation}

The second and third order terms can be obtained from
\tab{tab:HopfTable} by calculating $\vec{F_{xx}}$ and
$\vec{F_{xxx}}$ at the bifurcation point where $\nu = 1$ and
$\vec{x_s} = (1,1)^\trp$.  By definition
\begin{equation} 
  \label{eq:fxx}
  \Dfxx(\bxi,\bet) = \sum_{i,j=1}^{2}
    \left.\frac{\partial^2\vec{F}}
    {\partial x_i \partial x_j}
  \right|_{\stab{x}}\!\!
  \xi_i \eta_j \qquad \text{and} \qquad
  \Dfxxx(\bxi,\bet,\bzt) = \sum_{i,j,k=1}^{2}
  \left.\frac{\partial^3\vec{F}}
    {\partial x_i \partial x_j \partial x_k}
  \right|_{\stab{x}}\!\!
  \xi_i \eta_j \zeta_k.
\end{equation}
From \eqn{eq:Roessler} and \eqn{eq:fxx}, we get
\begin{align} 
  \left . \frac{\partial^2\vec{F}}{\partial x_1^2}
  \right|_{\stab{x}} &= 
  \begin{pmatrix}
    0 \\ 0
  \end{pmatrix}, \qquad
  \left . \frac{\partial^2\vec{F}}{\partial x_1 \partial x_2}
  \right|_{\stab{x}} = 
  \left . \frac{\partial^2\vec{F}}{\partial x_2 \partial x_1}
  \right|_{\stab{x}} = 
  \left . \frac{\partial^2\vec{F}}{\partial x_2^2}
  \right|_{\stab{x}} = 
  \begin{pmatrix}
    -2 \\ 2
  \end{pmatrix}.
  \label{eq:fxxs}
\end{align}

\noindent\textbf{Second order terms}\\
From \eqn{eq:fxx} and \eqn{eq:fxxs} we get
\begin{align}
  \BPhi_{200} &= \tfrac{1}{2} \vec{F_{xx}(u,u)} = (8 + 2\I,-8 - 2\I)^\trp,\\
  \BPhi_{110} &= \vec{F_{xx}(u,\overline u)} = (4,-4)^\trp.
\end{align}
The coefficients $\vec{h}_{200}$, $\vec{h}_{020}$, and
$\vec{h}_{110}$ can then be found by solving the linear
equations
\begin{align}
  (\vec{J} - 2\I\vec{I}) \mdot \vec{h}_{200} 
  &= -\BPhi_{200} = -(4 + 2\I,-4 - 2\I)^\trp,\\
  \vec{J} \mdot \vec{h}_{110} &= -\BPhi_{110} =-(4,-4)^\trp.
\end{align}
giving
\begin{align} 
  \label{eq:h200}
  \vec{h}_{200} &= \cc{\vec{h}}_{020} = -\frac{2}{3} (5\I,2 - 4\I)^\trp,\\
  \vec{h}_{110} &= (4, 0)^\trp. 
  \label{eq:h110}
\end{align}

\noindent\textbf{Third order terms}\\
From \eqn{eq:Roessler}, \eqn{eq:fxx},\eqn{eq:h200}, and
\eqn{eq:h110}, we get 
\begin{equation}
  \begin{split}
    &
    \Dfxx(\vec{u},\vec{h}_{200}) =
    -\frac{4}{3}(3 - 11\I) \vec{v}, \qquad
    \Dfxx(\vec{u},\vec{h}_{110}) =     
    -(2 + 6\I) \vec{v}, \qquad
    \Dfxx(\cc{\vec{u}},\vec{h}_{200}) =
    -\frac{4}{3}(1 - 7\I)\vec{v},\\
    &\Dfxxx(\vec{u},\vec{u},\vec{u}) = 24\I \vec{v}, \qquad
    \Dfxxx(\vec{u},\vec{u},\cc{\vec{u}}) = (16 + 8\I) \vec{v},
  \end{split}
\end{equation}
where $\vec{v} = (1,-1)^\trp$.  The third order coefficients
can then be found by solving the following linear
equations.  Solving
\begin{equation}
  (\vec{J} - 3 \I\vec{I}) \mdot \vec{h}_{300} 
  = -\BPhi_{300} = -\frac{4}{3}(3 - 14\I, -3 + 14\I)^\trp
  \label{eq:GSRes}
\end{equation}
gives
\begin{equation}
  \vec{h}_{300} = \cc{\vec{h}}_{030} = 
  -\frac{1}{6}(-45 + 5\I,42 + 9\I)^\trp.
\end{equation}
Solving
\begin{equation}
  \begin{split}
    (\vec{J}-\I\vec{I}) \mdot \vec{h}_{210} &= -\BPhi_{210} =
    -\frac{4}{3}(1 - 4\I,-1 + 4\I)^\trp,\\
    \vec{u}^*\mdot\vec{h}_{210} &= 0,
  \end{split}
\end{equation}
gives
\begin{equation} 
  \vec{h}_{210} = \cc{\vec{h}}_{120} = 
  \frac{1}{3}(3 + 5\I,-4 - \I).
\end{equation}

Using \tab{tab:HopfTable}, we may find the resonant
coefficient $g_3$ from \eqn{eq:GSRes} as
\begin{equation}
  g_3 = \vec{u}^* \mdot \BPhi_{210} = -(1 + \frac{5}{3}\I).
\end{equation}
Here we notice that the resonant coefficient satisfies $\re
g_3 < 0$ for all $\nu$ showing that the Hopf bifurcation is
always supercritical.

\vspace{2.5\parsep}
\noindent\textbf{Unfolding terms}\\
We first note that 
\begin{equation}
  \vec{F}_\nu = \left. \frac{\partial \vec{F}}{\partial\nu} 
  \right|_{\stab{x}} = (1,0)^\trp,
\end{equation}
implying that $\vec{h}_{001}$ can be found from the equation
\begin{equation}
  \vec{J} \mdot \vec{h}_{001} = -(1,0)^\trp,
\end{equation}
giving
\begin {equation}
  \vec{h}_{001} = (-1,1)^\trp.
\end{equation}
Observing that $\vec{F_{\vec{x}\nu} = 0}$, we may calculate
the coefficient $\sigma_1$ as
\begin{equation}
  \sigma_1 = \vec{u}^* \mdot \Dfxx (\vec{u},\vec{h}_{001}) =  -1+\I.
\end{equation}

Finally, we may find the coefficient vector $\vec{h}_{101}$ by
solving
\begin{equation}
  \begin{split}
    (\vec{J}-\I\vec{I}) \mdot \vec{h}_{101} &= (4,-4)^\trp,\\
    \vec{u}^* \mdot \vec{h}_{101} &= 0,
  \end{split}
\end{equation}
which gives
\begin{equation} 
  \vec{h}_{101} = \cc{\vec{h}}_{011} = (1 - \I,\I)^\trp.
\end{equation}

\noindent\textbf{Amplitude equation}\\
In conclusion, the amplitude equation \eqn{eq:HopfAmpl}
associated with the supercritical Hopf bifurcation in the
Gray-Scott model becomes
\begin{equation}
  \dot{z} =  \I z - (1 - \I)\nu z - (1 + \frac{5}{3}\I) z|z|^2.
\end{equation}

Any solution to the amplitude equation can then be
transformed back to the concentration space by using 
\begin{equation}
  \begin{split}
  &\vec{x} = 
  \begin{pmatrix}
    1 \\ 1
  \end{pmatrix} +
  \begin{pmatrix}
    -1 \\ 1    
  \end{pmatrix}\nu +
  \biggl[
  \begin{pmatrix}
    -2 \\ 1 + \I
  \end{pmatrix} z + \mathrm{c.c.}
  \biggr] -
  \frac{2}{3}
  \biggl[
  \begin{pmatrix}
    5\I \\ 2 - 4\I
  \end{pmatrix} z^2 + \mathrm{c.c.}
  \biggr] -
  \begin{pmatrix}
    4 \\ 0
  \end{pmatrix} \abs{z}^2 - \\
  &\qquad\qquad
  \frac{1}{6}
  \biggl[
  \begin{pmatrix}
    -45 + 5\I\\ 42 + 9\I
  \end{pmatrix} z^3 +
  2
  \begin{pmatrix}
    3 + 5\I\\ -4 - \I
  \end{pmatrix} z\abs{z}^2 + \mathrm{c.c.}
  \biggr] +
  \biggl[
  \begin{pmatrix}
    1 + \I\\ \I
  \end{pmatrix}\nu z + \mathrm{c.c.}
  \biggr],
  \end{split}
\end{equation}
where $\mathrm{c.c.}$ denotes the complex conjugate of all
terms within a particular square bracket.

\subsection{The Hénon map}
As an example of an iterated map which exhibits a period
doubling bifurcation, we consider the Hénon map
\cite{Henon76}
\begin{equation}
  \label{eq:Henon}
  \begin{pmatrix}
    x_1 \\
    x_2
  \end{pmatrix} \mapsto
  \begin{pmatrix}
    a - x_1^2 + bx_2\\
    x_1
  \end{pmatrix},
\end{equation}
and derive an explicit amplitude equation for it using the
parameter $a$ as bifurcation parameter.  We observe that
\eqn{eq:Henon} admits the non-trivial fixed point solution
\begin{equation}
  x_1 = x_2 = \frac{1}{2} \bigl(
  b - 1 + \sqrt{4a - (b-1)^2}
  \:\bigr),
\end{equation}
with corresponding Jacobian
\begin{equation}
  \vec{J}(a) = 
  \begin{pmatrix}
    1 - b - \sqrt{4a + (b-1)^2)} & b\\
                               1 & 0
  \end{pmatrix}.
\end{equation}
Clearly, $\vec{J}(a)$ has an eigenvalue $\lambda = -1$ if
\begin{equation}
  a = \frac{3(b-1)^2}{4}.
\end{equation}
The right and left eigenvectors associated with the critical
eigenvalue are
\begin{align}
  \vec{u}   &= (-1,1)^\trp,\\
  \vec{u}^* &= \frac{1}{1+b}(-1,b).
\end{align}
\noindent\textbf{Second and third order terms}\\
Observing that $\Dfxx(\bxi,\bet) = (-2 \xi_1 \eta_1, 0)^\trp$, we find
\begin{equation}
  \tfrac{1}{2} \Dfxx (\vec{u},\vec{u}) = (-1,0)^\trp.
\end{equation}
The second order coefficient vector $\vec{h}_{20}$ can then be
found from \tab{tab:DoubleTable} by solving the linear
equation
\begin{equation}
  (\vec{J} - \vec{I})\mdot \vec{h}_{20} = (1,0)^\trp,
\end{equation}
namely
\begin{equation}
  \vec{h}_{20} = \frac{1}{2(b-1)}(1,1)^\trp.
\end{equation}
Noting that $\Dfxxx = \vec{0}$, we use \tab{tab:DoubleTable}
and calculate the third order resonant term $g_3$ as
\begin{equation}
  g_3 = \vec{u}^* \mdot \Dfxx (\vec{u},\vec{h}_{20}) =
  \vec{u}^* \mdot\, (\frac{1}{b-1},0)^\trp = \frac{1}{1-b^2}.
\end{equation}

\noindent\textbf{Unfolding terms}\\
To find the unfolding of the amplitude equation associated
with the period doubling in \eqn{eq:Henon}, we first determine
the constant coefficient vector $\vec{h}_{01}$.  From
\tab{tab:DoubleTable}, we get
\begin{equation}
  (\vec{J} - \vec{I}) \mdot \vec{h}_{01} = -\vec{F}_a =
  -(1,0)^\trp
\end{equation}
which readily is solved for $\vec{h}_{01}$ as
\begin{equation}
  \vec{h}_{01} =
  \frac{1}{2(1-b)} (1,1)^\trp.
\end{equation}
Here we note that all elements of the matrix
$\vec{F}_{\vec{x}a}$ are zero.  Using \tab{tab:DoubleTable},
we conclude that the unfolding coefficient $\sigma_1$ becomes
\begin{equation}
  \sigma_1 = \vec{u}^* \mdot \Dfxx (\vec{u},\vec{h}_{01}) = 
  \frac{1}{b^2-1}.
\end{equation}
If we furthermore define the distance $\mu$ from the
bifurcation point as $\mu = a - \tfrac{3(1-b)^2}{4}$, we
arrive at the following amplitude equation for the period
doubling bifurcation in the Hénon map \eqn{eq:Henon}
\begin{equation}
  z \mapsto (\frac{1}{b^2-1}\mu - 1)z + \frac{1}{1 - b^2}z^3.
  \label{eq:HenonAmpl}
\end{equation}
For the doubly iterate of the amplitude equation, we find the
following two non-trivial fixed point solutions
\begin{equation}
  z_\pm = \pm\sqrt{\mu}.
\end{equation}
For the Hénon map, this corresponds to the emergence of a two
stable period-two orbits whose location in $(x_1,x_2)$-phase
space and parametric dependence on $\mu$ is determined via the
amplitude transformation from \tab{tab:Phys} as
\begin{align}
  \begin{pmatrix}
    x_1 \\
    x_2 
  \end{pmatrix}_{\!\!\pm} =
  \tfrac{1}{2}(1-b)
  \begin{pmatrix}
    1 \\
    1
  \end{pmatrix}\sqrt{\mu}
  \pm
  \begin{pmatrix}
    -1 \\
    1
  \end{pmatrix}\sqrt{\mu} +
  \frac{1}{2(b-1)}
  \begin{pmatrix}
    1 \\
    1 
  \end{pmatrix}\mu.
\end{align}

Just as in the Lorenz model \eqn{eq:Lorenz}, we have not shown
the calculation of the coefficient vectors $\vec{h}_{30}$ and
$\vec{h}_{11}$, since these accidentally cancel when inserted
into the amplitude transformation.

\newpage
\bibliography{dynamic}
\bibliographystyle{prsty}

\newpage
\noindent\Large\textbf{Captions of tables and figures}
\normalsize

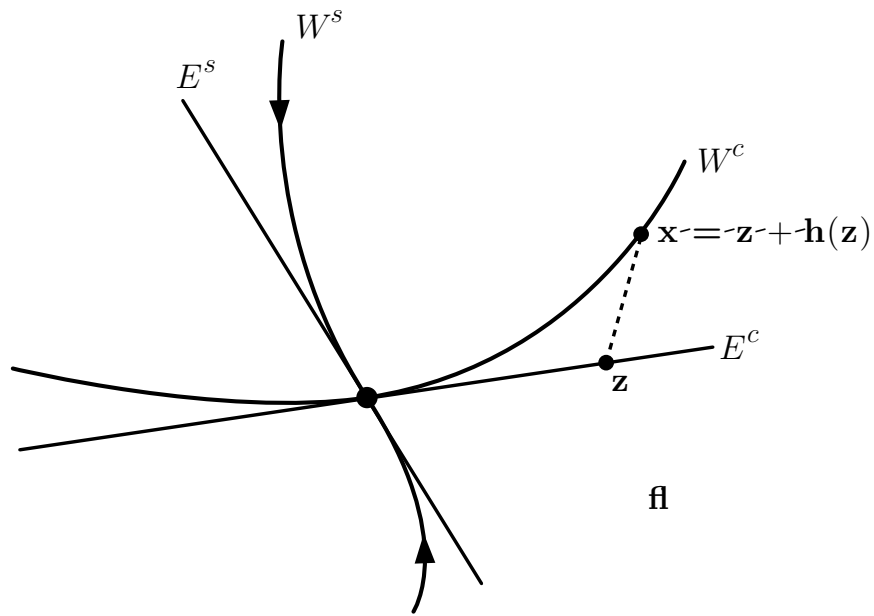
\begin{figure}[tb]
  \begin{center}
    \input{fig1.tex}
    \leavevmode
    \caption{}%
    \label{fig:AmpMap}
  \end{center}
\end{figure}

\vspace{2.5mm}%
\noindent\textbf{Figure~\ref{fig:AmpMap}}: Geometrical illustration of
the parameterization of the center manifold by
\eqn{eq:CenterParam}.  A point $\vec{z} \in \Ec$ is mapped by
the transformation $\vec{h}$ to the point $\vec{h}(\vec{z})$
so that $\vec{z} + \vec{h}(\vec{z}) \in \Wc$.

\vspace{2.5mm}%
\noindent\textbf{Table~\ref{tab:SaddleTable}}: Expressions for
calculating the coefficients of the amplitude transformation
and the amplitude equation for each of six bifurcations:
saddle-node, transcritical, and pitchfork bifurcations for
flows and iterated maps.  At the bifurcation, the Jacobian
$\vec{J}$ has right and left eigenvectors $\vec{u}$ and
$\vec{u}^*$ respectively corresponding to a critical
eigenvalue $\lambda=0$ for flows and $\lambda=1$ for iterated
maps.  The amplitude transformation $\vec{x} = \vec{z} +
\vec{h}(\vec{z},\mu)$, $\vec{z} = \vec{u}z$, transforms a
solution $z(t)$ of the amplitude equation to the motion
$\vec{x}(t)$ on the unfolded center manifold for the dynamical
system.  The vector coefficients $\vec{h}_{pq}$ written as
$\vec{h}_p$) are determined as solutions to the linear
equations indicated, in terms of the derivatives of the vector
field $\vec{F}$.  The coefficients of the amplitude equations
can then be found through the explicit expressions indicated,
in terms of $\vec{F}$ and $\vec{h}_{pq}$.

\vspace{2.5mm}%
\noindent\textbf{Table~\ref{tab:HopfTable}}: 
Formulae for calculating the coefficients of the amplitude
transformation and the amplitude equation for the Hopf
bifurcation and the Neimark-Sacker bifurcation.  At the
bifurcation, the Jacobian $\vec{J}$ has two complex conjugate
eigenvectors $\vec{u}$ and $\cc{\vec{u}}$ and left
eigenvectors $\vec{u}^*$ and $\cc{\vec{u}}^*$ corresponding to
critical eigenvalues $\lambda=\pm\I\omega_0$ for flows (Hopf
bifurcation) and $\lambda=\e^{\pm\I\!\theta_0}$ for iterated
maps (Neimark-Sacker bifurcation).  The amplitude
transformation $\vec{x} = \vec{z} + \vec{h}(\vec{z},\mu)$,
$\vec{z} = \vec{u}z + \cc{\vec{u}z}$, transforms a solution
$z(t)$ of the amplitude equation to the motion $\vec{x}(t)$ on
the unfolded center manifold for the dynamical system.  The
vector coefficients $\vec{h}_{ijk}$ are determined as
solutions to the linear equations indicated, in terms of the
derivatives of the vector field $\vec{F}$.  The coefficients
of the amplitude equations can then be found through the
explicit expressions indicated, in terms of $\vec{F}$ and
$\vec{h}_{ijk}$.

\vspace{2.5mm}%
\noindent\textbf{Table~\ref{tab:DoubleTable}}: 
Expressions for calculating the coefficients of the amplitude
transformation and the amplitude equation for the period
doubling bifurcation.  At the bifurcation, the Jacobian
$\vec{J}$ has right and left eigenvectors $\vec{u}$ and
$\vec{u}^*$ respectively corresponding to a critical
eigenvalue $\lambda = -1$.  The amplitude transformation
$\vec{z} + \vec{h}(\vec{z},\mu)$, $\vec{z} = \vec{u}z$,
transforms a solution $z(t)$ to the amplitude equation to the
motion $\vec{x}(t)$ on the unfolded center manifold for the
dynamical system.  The vector coefficients $\vec{h}_{pq}$ are
determined as solutions to the linear equations indicated, in
terms of the derivatives of the vector field $\vec{F}$.  The
coefficients of the amplitude equation can then be found
through the explicit expressions indicated, in terms of
$\vec{F}$ and $\vec{h}_{pq}$.

\vspace{2.5mm}%
\noindent\textbf{Table~\ref{tab:Phys}}: 
Table showing the limit set solutions (stationary points,
fixed points, or period orbits) arising from: a) saddle-node,
transcritical, and pitchfork bifurcations in flows and
iterated maps, b) Hopf bifurcations in flows, and c) period
doubling and Neimark-Sacker bifurcation in iterated maps.  The
second column shows the well-known solutions to the amplitude
equation, whereas column three gives the physical
representation of these in terms of the amplitude
transformation.  Since the period doubling in an iterated map
corresponds to a pitchfork bifurcation in the doubly iterated
map, the results for these two bifurcations are identical as
indicated in the table.  The solution shown for the Hopf and
Neimark-Sacker bifurcation corresponds to a periodic solution
for the Hopf bifurcation.  The limit cycle corresponds to the
invariant ellipse emerging at the Neimark-Sacker bifurcation.

\pagestyle{empty}
\clearpage
\renewcommand{\baselinestretch}{1.0}

\clearpage
\begin{sidewaystable}
  \begin{center}
    \newlength{\TopSpace}
\newlength{\BotSpace}
\newlength{\IntSpace}
\newlength{\AlgSpace}
\setlength{\TopSpace}{.0pt}
\setlength{\BotSpace}{.0pt}
\setlength{\IntSpace}{7.0pt}
\setlength{\AlgSpace}{-4.0pt}

\newlength{\TableLength}
\newlength{\TableLengthA}
\newlength{\TableLengthB}
\newlength{\TableLengthC}
\setlength{\TableLengthA}{4.1cm}
\setlength{\TableLengthB}{5.60cm}
\setlength{\TableLengthC}{6.87cm}

\newlength{\LinearLengthA}
\newlength{\LinearLengthB}
\newlength{\LinearLengthC}
\setlength{\LinearLengthA}{3.8cm}
\setlength{\LinearLengthB}{5.3cm}
\setlength{\LinearLengthC}{6.0cm}

\newcommand{\AlignKolon}[1]{%
  \hspace*{-6.0pt}\parbox{20pt}{\hspace*{\fill}$#1$:\hspace*{0.0pt}}}

\newcommand{\BoxLinA}{%
  \vspace*{\TopSpace}
  \AlignKolon{z^2}
  $\begin{eqalign}
    \J \mdot \vec{h}_{20} &=
    -\tfrac{1}{2}\vec{Q} \mdot \Dfxx (\vec{u},\vec{u})\\[\AlgSpace]
    \vec{u}^*\mdot \vec{h}_{20} &= 0
  \end{eqalign}$

  \vspace*{\IntSpace}
  \AlignKolon{\mu}
  $\begin{eqalign}
    \J \mdot \vec{h}_{01} &= -\vec{Q} \mdot \Dfp\\[\AlgSpace]
    \vec{u}^*\mdot \vec{h}_{01} &= 0
  \end{eqalign}$
  }

\newcommand{\BoxResA}{%
  \vspace*{\TopSpace}
  $\begin{eqalign}
    g_2 &= \tfrac{1}{2} \vec{u}^* \mdot \Dfxx (\vec{u},\vec{u})\\[\AlgSpace]
    \sigma_0 &= \vec{u}^* \mdot \Dfp
  \end{eqalign}$
  }

\newcommand{\BoxLinB}{%
  \vspace*{\TopSpace}
  \AlignKolon{z^2}
  $\begin{eqalign}
    \J \mdot \vec{h}_{20} &= 
    -\tfrac{1}{2}\vec{Q} \mdot \Dfxx (\vec{u},\vec{u})\\[\AlgSpace]
    \vec{u}^*\mdot \vec{h}_{20} &= 0
  \end{eqalign}$

  \vspace*{\IntSpace}
  \AlignKolon{\mu}
  $\begin{eqalign}
    \J \mdot \vec{h}_{01} &= -\Dfp\\[\AlgSpace]
    \vec{u}^* \mdot \vec{h}_{01} &= 0
  \end{eqalign}$

  \vspace*{\IntSpace}
  \AlignKolon{\mu z}
  $\begin{eqalign}
    \J \mdot \vec{h}_{11} &= -\vec{Q} \mdot \bigl(\Dfxp \mdot \vec{u} + 
    \Dfxx (\vec{u},\vec{h}_{01})\bigr)\\[\AlgSpace]
    \vec{u}^* \mdot \vec{h}_{11} &= 0
  \end{eqalign}$
  }

\newcommand{\BoxResB}{%
  \vspace*{\TopSpace}
  $\begin{eqalign}
    g_2 &= \tfrac{1}{2} \vec{u}^* \mdot \Dfxx (\vec{u},\vec{u})\\[\AlgSpace]
    \sigma_1 &= \vec{u}^* \mdot \Dfxp \mdot \vec{u} + 
    \vec{u}^* \mdot \Dfxx (\vec{u},\vec{h}_{01})
  \end{eqalign}$
  }

\newcommand{\BoxLinC}{%
  \vspace*{\TopSpace}
  \AlignKolon{z^2}
  $\begin{eqalign}
    \J \mdot \vec{h}_{20} &= 
    -\tfrac{1}{2}\vec{Q} \mdot \Dfxx (\vec{u},\vec{u})\\[\AlgSpace]
    \vec{u}^*\mdot \vec{h}_{20} &= 0
  \end{eqalign}$

  \vspace*{\IntSpace}
  \AlignKolon{z^3}
  $\begin{eqalign}
    \J \mdot \vec{h}_{30} &= 
    -\vec{Q} \mdot \Dfxx (\vec{u},\vec{h}_{20}) +
    \tfrac{1}{6} \vec{Q} \mdot \Dfxxx (\vec{u},\vec{u},\vec{u})\\[\AlgSpace]
    \vec{u}^* \mdot \vec{h}_{30} &= 0
  \end{eqalign}$

  \vspace*{\IntSpace}
  \AlignKolon{\mu}
  $\begin{eqalign}
    \J \mdot \vec{h}_{01} &= -\Dfp\\[\AlgSpace]
    \vec{u}^* \mdot \vec{h}_{01} &= 0
  \end{eqalign}$

  \vspace*{\IntSpace}
  \AlignKolon{\mu z}
  $\begin{eqalign}
    \J \mdot \vec{h}_{11} &= -\vec{Q} \mdot \bigl(\Dfxp \mdot \vec{u} + 
    \Dfxx (\vec{u},\vec{h}_{01})\bigr)\\[\AlgSpace]
    \vec{u}^* \mdot \vec{h}_{11} &= 0
  \end{eqalign}$
  }

\newcommand{\BoxResC}{%
  \vspace*{\TopSpace}
  $\begin{eqalign}
    g_3 &= \vec{u}^* \mdot \Dfxx (\vec{u},\vec{h}_{20}) + 
    \tfrac{1}{6}\vec{u}^* \mdot \Dfxxx (\vec{u},\vec{u},\vec{u})\\[\AlgSpace]
    \sigma_1 &= \vec{u}^* \mdot \Dfxp \mdot \vec{u} + 
    \vec{u}^* \mdot \Dfxx (\vec{u},\vec{h}_{01})
  \end{eqalign}$
  }

\begin{tabular}{%
  |>{\footnotesize}m{2.7cm}<{}
  |>{\footnotesize\vspace{1.0mm}} m{\TableLengthA} <{\vspace{1.0mm}}
  |>{\footnotesize\vspace{1.0mm}} m{\TableLengthB} <{\vspace{1.0mm}}
  |>{\footnotesize\vspace{1.0mm}} m{\TableLengthC} <{\vspace{1.0mm}}
  |}\hline
\textsf{Bifurcation} 
& 
\multicolumn{1}{c|}{\emph{\footnotesize Saddle-node}} 
& 
\multicolumn{1}{c|}{\emph{\footnotesize Transcritical}} 
& 
\multicolumn{1}{c|}{\emph{\footnotesize Pitchfork}} 
\\\hline\hline
\textsf{Transformation}\newline
$\vec{x} = \vec{z} + \vec{h}(\vec{z},\mu)$
& $\vec{x} = \vec{u} z + \vec{h}_{20} z^2 + \vec{h}_{01}\mu$
& $\vec{x} = \vec{u} z + \vec{h}_{20} z^2 + \vec{h}_{01}\mu + \vec{h}_{11}z\mu$
& $\vec{x} = \vec{u} z + \vec{h}_{20} z^2 + \vec{h}_{30} z^3 + 
\vec{h}_{01}\mu + \vec{h}_{11}z\mu$
\\\hline
\textsf{Linear equations}\newline 
\textsf{for $\vec{h}_{pq}$} 
& \vspace*{-59.5pt}\BoxLinA
& \vspace*{-30.5pt}\BoxLinB
& \BoxLinC
\\\hline
\textsf{Resonant coefficients} 
& \BoxResA
& \BoxResB
& \BoxResC
\\\hline
\textsf{Amplitude equation} 
& $\dot{z} = \sigma_0\mu + g_2 z^2$\newline
  $z \mapsto \sigma_0\mu + z + g_2 z^2$
& $\dot{z} = \sigma_1\mu z + g_2 z^2$\newline
  $z \mapsto (1+\sigma_1\mu)z + g_2 z^2$
& $\dot{z} = \sigma_1\mu z + g_3 z^3$\newline
  $z \mapsto (1+\sigma_1\mu)z + g_3 z^3$
\\\hline
\multicolumn{4}{|>{\footnotesize\vspace{1.0mm}} m{8.0cm} 
  <{\vspace{1.0mm}}|}{\footnotesize%
  $\begin{aligned}
    \J &= \vec{J} \quad\text{for flows},\\
    \J &= \vec{J}-\vec{I} \quad\text{for iterated maps},\\
    \vec{Q} \mdot \vec{x} &= \vec{x} - (\vec{u}^* \mdot \vec{x})\vec{u}.
  \end{aligned}$
  }%
  \\\hline
\end{tabular}


    \caption{%
      }%
    \label{tab:SaddleTable}
  \end{center}
\end{sidewaystable}

\clearpage
\begin{sidewaystable}[t!]
  \begin{center}
    \setlength{\TopSpace}{.0pt}
\setlength{\BotSpace}{.0pt}
\setlength{\IntSpace}{9.0pt}
\setlength{\AlgSpace}{-5.0pt}

\setlength{\TableLengthA}{10.4cm}
\setlength{\LinearLengthA}{6.0cm}
\newcommand{\AlignKolon}[1]{%
  \hspace*{-6.0pt}\parbox{22pt}{\hspace*{\fill}$#1$:\hspace*{0.0pt}}}

\newcommand{\BoxLinA}{%
  \vspace*{\TopSpace}
  \AlignKolon{z^2}
  $\begin{eqalign}
    (\vec{J} - \gamma_2\vec{I}) \mdot \vec{h}_{200} &= 
    -\tfrac{1}{2}\Dfxx ({\vec{u},\vec{u}})\\[\AlgSpace]
    \vec{h}_{020} &= \cc{\vec{h}}_{200}
  \end{eqalign}$

  \vspace*{\IntSpace}
  \AlignKolon{\abs{z}^2}
  $\begin{eqalign}
    (\vec{J} - \gamma_0\vec{I}) \mdot \vec{h}_{110} &= 
    -\Dfxx ({\vec{u},\cc{\vec{u}}})
  \end{eqalign}$

  \vspace*{\IntSpace}
  \AlignKolon{z^3}
  $\begin{eqalign}
    (\vec{J} - \gamma_3\vec{I}) \mdot 
    \vec{h}_{300} &= -\Dfxx (\vec{u},\vec{h}_{200}) -
    \tfrac{1}{6}\Dfxxx (\vec{u},\vec{u}, \vec{u}),\\[\AlgSpace]
    \vec{h}_{030} &= \cc{\vec{h}}_{300}
  \end{eqalign}$

  \vspace*{\IntSpace}
  \AlignKolon{z\abs{z}^2}
  $\begin{eqalign}
    (\vec{J} - \gamma_1 \vec{I}) \mdot 
    \vec{h}_{210} &= -\vec{Q} \mdot \bigl( \Dfxx (\vec{u},\vec{h}_{110}) + 
    \Dfxx (\cc{\vec{u}},\vec{h}_{200}) + 
    \tfrac{1}{2}\Dfxxx (\vec{u},\vec{u},\cc{\vec{u}}) \bigl)\\[\AlgSpace]
    \vec{u}^* \mdot \vec{h}_{210} &= 0\\[\AlgSpace]
    \vec{h}_{120} &= \cc{\vec{h}}_{210}
  \end{eqalign}$

  \vspace*{\IntSpace}
  \AlignKolon{\mu}
  $\begin{eqalign}
    (\vec{J} - \gamma_0\vec{I})\mdot \vec{h}_{001} &= -\Dfp
  \end{eqalign}$

  \vspace*{\IntSpace}
  \AlignKolon{\mu z}
  $\begin{eqalign}
    (\vec{J} - \gamma_1\vec{I}) \mdot \vec{h}_{101} &= 
    -\vec{Q} \mdot \bigl(\Dfxp \mdot \vec{u} + 
    \Dfxx (\vec{u},\vec{h}_{001}) \bigr)\\[\AlgSpace]
    \vec{u}^* \mdot \vec{h}_{101} &= 0\\[\AlgSpace]
    \vec{h}_{011} &= \cc{\vec{h}}_{101}
  \end{eqalign}$
  }

\newcommand{\BoxResA}{%
  \vspace*{\TopSpace}
  $\begin{eqalign}
    g_3 &= \vec{u}^* \mdot \Dfxx (\vec{u},\vec{h}_{110})
    + \vec{u}^* \mdot \Dfxx (\cc{\vec{u}},\vec{h}_{200}) + 
    \tfrac{1}{2}\vec{u}^* \mdot \Dfxxx(\vec{u},\vec{u},\cc{\vec{u}})
    \\[\AlgSpace]
    \sigma_1 &= \vec{u}^* \mdot \Dfxp \mdot \vec{u} + 
    \vec{u}^* \mdot \Dfxx (\vec{u},\vec{h}_{001})
  \end{eqalign}$
  }

\begin{tabular}{%
  |>{\footnotesize} m{2.7cm} <{}
  |>{\footnotesize\vspace{1.0mm}} m{\TableLengthA} <{\vspace{1.0mm}}
  |}\hline
\multicolumn{2}{|c|}{\emph{\footnotesize Hopf and
    Neimark-Sacker Bifurcations}} 
\\\hline\hline
\textsf{Transformation}\newline
$\vec{x} = \vec{z} + \vec{h}(\vec{z},\mu)$ &
$\begin{aligned}
\vec{x} &= 
\vec{u}z + \cc{\vec{u}z} +
\vec{h}_{200}z^2 + \vec{h}_{110}\abs{z}^2  + \vec{h}_{020}\cc{z}^2 +
\vec{h}_{300}z^3 + \vec{h}_{210}\abs{z}^2z +\\
&\qquad\vec{h}_{120}\abs{z}^2 \cc{z}  + \vec{h}_{030}\cc{z}^3  +
\vec{h}_{001}\mu + (\vec{h}_{101}z + \vec{h}_{011}\cc{z})\mu
\end{aligned}$
\\\hline
\textsf{Linear equations}\newline
\textsf{for $\vec{h}_{ijk}$} 
& \BoxLinA
\\\hline
\textsf{Resonant coefficients} 
& \BoxResA
\\\hline
\textsf{Amplitude equation} 
& $\dot{z} = (\I\omega_0 + \sigma_1 \mu)z + g_3 z\abs{z}^2$\newline
  $z \mapsto (\e^{\I\! \theta_0} + \sigma_1 \mu)z + g_3 z\abs{z}^2$
\\\hline
\multicolumn{2}{|>{\footnotesize\vspace{1.0mm}} m{8.0cm} <{\vspace{1.0mm}}|}{%
  $\begin{aligned}
    \gamma_k &= k\I\omega_0 \quad\text{for flows},\\
    \gamma_k &= \e^{k\!\I\!\theta_0} \quad\text{for iterated maps},\\
    \vec{Q} \mdot \vec{x} &= 
    \vec{x} - (\vec{u}^* \mdot \vec{x})\vec{u}.
  \end{aligned}$
  }%
  \\\hline
\end{tabular}


    \caption{%
      }%
    \label{tab:HopfTable}
  \end{center}
\end{sidewaystable}

\clearpage
\begin{table}
  \begin{center}
    \setlength{\TopSpace}{.0pt}
\setlength{\BotSpace}{.0pt}
\setlength{\IntSpace}{9.0pt}
\setlength{\AlgSpace}{-3.0pt}

\setlength{\TableLengthA}{7.5cm}
\setlength{\LinearLengthA}{6.0cm}
\newcommand{\AlignKolon}[1]{%
  \hspace*{0.0pt}\parbox{16pt}{\hspace*{\fill}$#1$:\hspace*{4.0pt}}}

\newcommand{\BoxLinA}{%
  \vspace*{\TopSpace}
  \AlignKolon{z^2}
  $\begin{eqalign}
    (\vec{J} - \vec{I}) \mdot \vec{h}_{20} &= 
    -\tfrac{1}{2} \Dfxx (\vec{u},\vec{u})
  \end{eqalign}$

  \vspace*{\IntSpace}
  \AlignKolon{z^3}
  $\begin{eqalign}
    (\vec{J} + \vec{I}) \mdot \vec{h}_{30} &=  
    -\vec{Q} \mdot \bigl( \Dfxx (\vec{u},\vec{h}_{20}) +
    \tfrac{1}{6}\Dfxxx (\vec{u},\vec{u},\vec{u})\bigr)\\[\AlgSpace]
    \vec{u}^* \mdot \vec{h}_{30} &= 0
  \end{eqalign}$

  \vspace*{\IntSpace}
  \AlignKolon{\mu}
  $\begin{eqalign}
    (\vec{J} - \vec{I})\mdot \vec{h}_{01} &= -\Dfp
  \end{eqalign}$

  \vspace*{\IntSpace}
  \AlignKolon{\mu z}
  $\begin{eqalign}
    (\vec{J} + \vec{I}) \mdot \vec{h}_{11} &= 
    -\vec{Q} \mdot \bigl( \vec{F}_{\vec{x}\mu} \mdot \vec{u} + 
    \Dfxx (\vec{u},\vec{h}_{01}) \bigr)\\[\AlgSpace]
    \vec{u}^* \mdot \vec{h}_{11} &= 0\\[\AlgSpace]
  \end{eqalign}$
  }

\newcommand{\BoxResA}{%
  \vspace*{\TopSpace}
  $\begin{eqalign}
    g_3 &= \vec{u}^* \mdot \Dfxx (\vec{u},\vec{h}_{20}) + 
    \tfrac{1}{6} \vec{u}^* \mdot \Dfxxx (\vec{u},\vec{u},\vec{u})
    \\[\AlgSpace]
    \sigma_1 &= \vec{u}^* \mdot \Dfxp \mdot \vec{u} + 
    \vec{u}^* \mdot \Dfxx (\vec{u},\vec{h}_{01})
  \end{eqalign}$
  }

\begin{tabular}{%
  |>{\footnotesize}m{2.7cm}<{}
  |>{\footnotesize\vspace{1.0mm}} m{\TableLengthA} <{\vspace{1.0mm}}
  |}\hline
\multicolumn{2}{|c|}{\emph{\footnotesize Period Doubling Bifurcation}} 
\\\hline\hline
\textsf{Transformation}\newline
$\vec{x} = \vec{z} + \vec{h}(\vec{z},\mu)$
& $\vec{x} = \vec{u} z + \vec{h}_{20} z^2 + \vec{h}_{30} z^3 + 
\vec{h}_{01}\mu + \vec{h}_{11}\mu z$
\\\hline
\textsf{Linear equations}\newline 
\textsf{for $\vec{h}_{pq}$} 
& \BoxLinA
\\\hline
\textsf{Resonant coefficients} 
& \BoxResA
\\\hline
\textsf{Amplitude equation} 
& $z \mapsto (\sigma_1 \mu - 1)z + g_3 z^3$
\\\hline
\multicolumn{2}{|>{\footnotesize\vspace{1.0mm}} m{8.0cm} <{\vspace{1.0mm}}|}{%
  $\begin{aligned}
    \vec{Q} \mdot \vec{x} &= 
    \vec{x} - (\vec{u}^* \mdot \vec{x})\vec{u}.
  \end{aligned}$
  }%
  \\\hline
\end{tabular}


    \caption{%
      }%
    \label{tab:DoubleTable}
  \end{center}
\end{table}

\begin{sidewaystable}
  \begin{center}
    \setlength{\TopSpace}{6.0pt}
\setlength{\BotSpace}{-6.0pt}
\setlength{\IntSpace}{9.0pt}
\setlength{\AlgSpace}{-3.0pt}

\setlength{\TableLengthA}{2.5cm}
\setlength{\LinearLengthA}{6.0cm}
\newcommand{\AlignKolon}[1]{%
  \hspace*{0.0pt}\parbox{16pt}{\hspace*{\fill}$#1$:\hspace*{4.0pt}}}

\newcommand{\BoxA}{%
  $\begin{eqalign}
    z_\pm = \pm a, \quad a = \sqrt{-\frac{\sigma_0\mu}{g_2}}
  \end{eqalign}$ 
  }

\newcommand{\BoxPhysA}{%
  $\begin{eqalign}
    \vec{x}_\pm = 
    \pm\vec{u} a + \vec{h}_{20} a^2 + \vec{h}_{01}\mu
  \end{eqalign}$ 
  }

\newcommand{\BoxB}{%
  $\begin{aligned}
    z_+ &= a, \quad a = -\frac{\sigma_1\mu}{g_2}\\
    z_0 &= 0
  \end{aligned}$ 
  }

\newcommand{\BoxPhysB}{%
  $\begin{aligned}
    \vec{x}_+ &= \vec{u} a + \vec{h}_{20} a^2 + \vec{h}_{01}\mu + 
    \vec{h}_{11} a\mu\\
    \vec{x}_0 &= \vec{h}_{01}\mu
  \end{aligned}$ 
  }

\newcommand{\BoxC}{%
  $\begin{aligned}
    z_\pm &= \pm a, \quad a = \sqrt{-\frac{\sigma_1\mu}{g_3}}\\
      z_0 &= 0
  \end{aligned}$ 
  }

\newcommand{\BoxPhysC}{%
  $\begin{aligned}
    \vec{x}_\pm &= 
    \pm\vec{u} a + \vec{h}_{20} a^2 \pm \vec{h}_{30} a^3 + 
    \vec{h}_{01}\mu \pm\vec{h}_{11}a\mu\\
    \vec{x}_0 &= \vec{h}_{01}\mu
  \end{aligned}$ 
  }

\newcommand{\BoxD}{%
  $\begin{aligned}
    z &= a \e^{\I\omega t}, \quad z_0 = 0,\\
    a &= \sqrt{-\sigma_1'\mu/g_3'},\\
    \omega &= \omega_0 + 
    (\sigma_1'' - \frac{g_3''}{g_3'}\sigma_1')\mu,\\
    g_3 &= g_3' + \I g_3'', \quad\sigma_1 = \sigma_1' + \I \sigma_1''
  \end{aligned}$
  }

\newcommand{\BoxPhysD}{%
  $\vec{x} = \bigl[
         \vec{u} a\e^{\I\omega t} + 
  \vec{h}_{200} a^2\e^{2\I\omega t} +
  \vec{h}_{300} a^3\e^{3\I\omega t} + $\newline
  \hspace*{1.0cm}$\vec{h}_{210} a^3\e^{\I\omega t} +
  \vec{h}_{101}a\e^{\I\omega t}\mu + \mathrm{c.c.}\bigr] +$\newline 
  \hspace*{1.3cm}$\vec{h}_{110} a^2 + \vec{h}_{001}\mu,$\newline
  $\vec{x}_0 = \vec{h}_{001}\mu$
  }

\newcommand{\BoxE}{%
  $\begin{aligned}
    z_\pm &= \pm a, \quad a = \sqrt{-\frac{\sigma_1\mu}{g_3}}\\
      z_0 &= 0
  \end{aligned}$ 
  }

\newcommand{\BoxPhysE}{%
  $\begin{aligned}
    \vec{x}_\pm &= 
    \pm\vec{u} a + \vec{h}_{20} a^2 \pm \vec{h}_{30} a^3 + 
    \vec{h}_{01}\mu \pm\vec{h}_{11}a\mu\\
    \vec{x}_0 &= \vec{h}_{01}\mu
  \end{aligned}$ 
  }

\begin{tabular}{%
  |>{\normalsize} m{2.5cm} <{}
  |>{\normalsize} m{4.8cm} <{}
  |>{\normalsize} m{6.8cm} <{}
  |>{\normalsize} m{5.0cm} <{}
  |}\cline{2-4}
\multicolumn{1}{c|}{} & 
\multicolumn{3}{c|}{\emph{\normalsize Limit set solutions}}
\\\hline
\multicolumn{1}{|c|}{\emph{\normalsize Bifurcation}} &  
 \multicolumn{1}{c|}{\emph{\normalsize Amplitude space}} & 
 \multicolumn{1}{c|}{\emph{\normalsize Physical space}} & 
 \multicolumn{1}{c|}{\emph{\normalsize Solution diagram}}
\\\hline\hline
\textsf{Saddle-node} 
& \BoxA
& \BoxPhysA
& \vspace*{\TopSpace}
\begin{center}
  \includegraphics[width=4.5cm]{saddle.ai}

  \vspace{-3.0mm}
\end{center}
\\\hline
\textsf{Transcritical}
& \BoxB
& \BoxPhysB
& \vspace*{\TopSpace}
\begin{center}
  \includegraphics[width=4.5cm]{trans.ai}

  \vspace{-3.0mm}
\end{center}
\\\hline
\textsf{Pitchfork/}\newline
\textsf{Period doubling}
& \BoxC
& \BoxPhysC
& \vspace*{\TopSpace}
\begin{center}
  \includegraphics[width=4.5cm]{pitch.ai}

  \vspace{-3.0mm}
\end{center}
\\\hline
\textsf{Hopf/Neimark}
& \BoxD
& \BoxPhysD
& \vspace*{\TopSpace}
\begin{center}
  \includegraphics[width=4.5cm]{hopf.ai}

  \vspace{-3.0mm}
\end{center}

\vspace*{\BotSpace}
\\\hline
\end{tabular}


    \caption{%
      }%
    \label{tab:Phys}
  \end{center}
\end{sidewaystable}


\end{document}